\documentclass[a4paper]{spie}  %>>> use for US letter paper
\usepackage{graphicx}
\usepackage[utf8]{inputenc}
\usepackage{url}
\usepackage{multirow}
\usepackage{array}
\usepackage{longtable}
\usepackage{makecell}

\title{Unveiling new stellar companions \\ from the PIONIER exozodi survey} 

\author{L. Marion\supit{a}, O. Absil\supit{a}, S. Ertel\supit{b,c}, J.-B. Lebouquin\supit{c}, D. Defr\`ere\supit{d}
\skiplinehalf
\supit{a}D\'epartement d'Astrophysique, G\'eophysique et Oc\'eanographie, Universit\'e de Li\`ege, All\'ee du Six Ao\^ut 17, 4000 Li\`ege, Belgium; \\
\supit{b}European Southern Observatory, Alonso de C\'ordova 3107, Vitacura, Santiago, Chile;\\
\supit{c}UJF-Grenoble 1 / CNRS-INSU, Institut de Plan\'etologie et d'Astrophysique de Grenoble (IPAG) UMR 5274, Grenoble, F-38041,
France;\\
\supit{d}Department of Astronomy, University of Arizona, 993 N. Cherry Ave, Tucson, AZ 85721, USA
}

\authorinfo{Send correspondence to L. Marion (email: marion@astro.ulg.ac.be, telephone: +32 (0)4 366 9758).}
\begin{document} 
  \maketitle 

%%%%%%%%%%%%%%%%%%%%%%%%%%%%%%%%%%%%%%%%%%%%%%%%%%%%%%%%%%%%% 
\begin{abstract}

The main goal of the EXOZODI survey is to detect and characterize circumstellar dust and to propose the first statistical study of exozodiacal disks in the near-infrared using telescopes in both hemispheres (VLTI and CHARA).  For this purpose, Ertel et al.\ have conducted in 2012 a survey of nearby main sequence stars with VLTI/PIONIER to search for the presence of circumstellar dust. This survey, carried out during 12 nights, comprises about 100 stars. For each star, we obtained typically three OBs and we searched for circumstellar emission based on the measurement of squared visibilities at short baselines. A drop in the measured visibilities with respect to the expected photospheric visibility indicates the presence of resolved emission around the target star. It is however generally not possible to conclude on the morphology of the detected emission based solely on the squared visibilities. Here, we focus on closure phases to search systematically for faint companions around the whole sample. Indeed, to derive robust statistics on the occurrence rate of bright exozodiacal disks, we need to discriminate between companions and disks. For this reason, the main goal of this paper is to discriminate between circumstellar disks (which show no closure phase provided that they are point-symmetric) and faint companions (point-like sources, creating non-zero closure phases). We also aim to reveal new companions that do not necessarily produce a significant signature in the squared visibilities, as the signature of the companion may show up more prominently in the closure phases. In this process, we reveal four new stellar companions with contrasts ranging from 2\% to 95\% (i.e., up to near-equal flux binaries). We also tentatively detect faint companions around one other target that will require follow-up observations to be confirmed or infirmed. We discuss the implications of these discoveries on the results of the exozodi survey.

\end{abstract}

%>>>> Include a list of keywords after the abstract 


%%%%%%%%%%%%%%%%%%%%%%%%%%%%%%%%%%%%%%%%%%%%%%%%%%%%%%%%%%%%%
\section{INTRODUCTION}
\label{sec:intro}  % \label{} allows reference to this section

Detecting faint companions close to bright stars is a real challenge due to the high contrast between the companion and the primary star and to the small angular separation between the two. Furthermore, the observations can be hampered by the presence of dust in the neighborhood of the companion that partially or totally hide it. However, the presence of dust could also reveal the planet thanks to gravitationnal interactions between them. It is thus of great interest to characterize this dust, and this is the purpose of the EXOZODI survey \cite{Ertel14}. Conversely, the observed (when observed) infrared excess caused by a companion could simulate the one caused by a disk and this is the main reason why searching for companion within the EXOZODI survey: we want to be sure that the observed infrared excess is caused by an extended source.

As it is now well known, the most luminous component after the star in a planetary system is generally the warm dust that reflects the starlight and emits its own radiation in the infrared. This dust (also called exozodiacal dust) is the subject of the EXOZODI project that has several goals: (i) carry out the first statistical near-IR interferometric survey of exozodiacal disks in both hemispheres thanks to one instrument in California (FLUOR at CHARA), and another in Chile (PIONIER at the VLTI), (ii) complete among the first dynamical simulations for the detected exozodis, thereby better assessing the origin of the exozodiacal disks, and in particular, their connection with the presence of planets and/or of an external Kuiper-like belt, (iii) realize a qualitative step forward in debris disks modeling, by merging the statistical and dynamical approaches and develop the first models able to investigate the coupled effects of collisions and dynamics\footnote{\url{http://ipag.osug.fr/~augereau/Site/ANR_EXOZODI.html}}.

The main goal of this paper is to search for companions within the EXOZODI sample that was observed in 2012 using PIONIER at VLTI (92 stars). The stars of this sample were especially chosen not to harbor any (sub-)stellar companion based on radial velocity measurements or direct imaging. However, these techniques are not able to detect everything. This is the reason why, here, we systematically search for the signature of faint companions in the EXOZODI interferometric data set. In Sect.~\ref{sec:search}, we present the original search method proposed by Absil et al.\ in 2011 \cite{Absil11} and how we have adapted it to this large sample. Sect.~\ref{sec:results} summarizes the results of the search, with four new stellar companion unambiguously detected and another potential detection. In Sect.~\ref{sec:sensitivity}, we discuss the PIONIER sensitivity to faint companions based on our sample.

%%-----------------------------------------------------------
\section{SEARCHING FOR COMPANIONS}  \label{sec:search}

In this section, we discuss how we search for companion based on our VLTI/PIONIER data and how we compute confidence
levels for detections. We first describe the initial method developped by Absil et al.\cite{Absil11}. Then, we explain how we modify this technique and why. 

In the paper of Absil et al.\cite{Absil11}, the search for companions is only based on the closure phases. Indeed, the closure phase (CP) is known to be sensitive to asymetries. So, if the observed structure is perfectly symetric, the closure phase will be strictly zero (possibly a constant offset from zero due to imperfect CP calibration) while the presence of a companion will create asymetries that will result in a non zero closure phase with variations when plotted as a function of the wavelength. Then, Absil et al.\cite{Absil11} define a field of view and compute a binary model considering the primary star at the center of the search region with an off-axis companion of various contrast $r$ starting at $r=0$, at each point $(x,y)$ of the field of view in the search region. After that, they compute the CP for each model. Then, they compare their model to the observations in terms of closure phase using a chi square test:
\[\chi^2 = \sum_{\lambda}\frac{(CP_{mod} - CP_{data})^2}{\sigma^2_{CP}}\]

However, based on our EXOZODI data set, we noticed that the use of the sole closure phase leads to a lot of false positive detections. Basically, we obtained many results (12 stars out of 92) with a significance level of detection between 3 and $4\sigma$, and it was not obvious to discriminate whether this was a true detection or not. As a consequence, we use the closure phases (CP) and the square visibilities in a combined way so that only the true companions, that show both a closure phase and a square visibility signature (at the same position in the search region), show a significant signature. Indeed, the square visibility will also have variations if a companion exists. However, the square visibilities are also sensitive to centrosymmetric circumstellar disks, which create a drop in visibility at all baselines \cite{2007A&A...475..243D}. As in [\citenum{Absil11}], we assume that both the primary and the secondary are unresolved, and we compute a binary model considering the primary star at the center of the search region with an off-axis companion of various contrast $r$ starting at $r<0$ (the reason of a negative contrast will be explained later in this section), at each point $(x,y)$ of the field of view in the search region. Then, we compute the CP and the square visibility for each model, and, we compare our model to the observations in terms of closure phase and square visibility using a combined chi square test:

\[\chi^2_{tot} = \underbrace{\sum\frac{(V^2_{mod} - V^2_{data})^2}{\sigma^2_{V^2}}}_{\chi^2_{V^2}} + \underbrace{\sum\frac{(CP_{mod} - CP_{data})^2}{\sigma^2_{CP}}}_{\chi^2_{CP}}\]

We do this for each model and obtain a chi square cube depending on the position in the field of view and on the contrast of the secondary relative to the primary: $\chi^2(x,y,r)$. Finally, we find the minimum $\chi^2$ as a function of the contrast for each position in the field of view, giving us chi square maps. This allows us to see the position in the field of view that correspond to the best contrast minimizing the chi square. 

This method is more robust since we need to have a signature in both the closure phases and the square visibilities to detect a companion. However, sometimes, the signature of a disk in the square visibility is so strong that the combined chi square is relatively high. To discriminate this kind of situation, we can inspect the chi squares maps for the squared visibilities and the closure phases separately. Indeed, if a companion is detected, its signature will be seen in both the closure phase and the square visibility, while a disk will only be seen in the square visibility (assuming a symmetric one). Furthermore, a small offset in the closure phase could simulate a companion (false positive detection). In this case, we won't see anything in the square visibility. This is the reason why, when we had a tentative detection in the combined $\chi^2$, we also had a look at the $\chi^2$ associated to the closure phase and the square visibility separately as well as at the fit of the closure phase and the fit of the square visibility. 

There are two main limitations to the detection of faint companions: the field of view and the maximum contrast reachable between the primary and the secondary star. Consequently, we first need to define a search region given by the maximum angular separation at which a companion can be found. This region is limited, as explained in [\citenum{Absil11}], by three factors: 

\begin{enumerate}
\item the efficiency of the single mode fiber. As explained in [\citenum{Absil11}], we can consider that the transmission profile of the fiber has a gaussian profile whose full width at half maximum (FWHM) is estimated to be 420 mas;
\item the optical path difference between the two fringe packets. To properly detect a companion using the standard PIONIER DRS, we need the two fringe packets associated to the companion and the star, to at least partially overlap. Furthermore, the two fringe packets need to be in the same scan. The smallest scan of PIONIER is about 60 $\mu$m and we estimate the size of one fringe packet to be $\lambda^2/\Delta\lambda\approx 27\mu$m when the H band signal is dispersed onto three spectral channel (the case here). This means that, in order to have the two fringe packets in the same scan and to partially overlap, the maximum optical path delay (OPD) we can admit is 27$\mu$m. The OPD separation of the fringe packets is given by $\Delta {\rm OPD} = B \Delta\theta \cos \theta$. Considering $\Delta {\rm OPD}_{\rm max}=27\mu$m, a mean projected baseline of $B = 20$ m, we find a $\Delta \theta_{\rm max}\approx 200$ mas, but if we consider the maximum baseline ($B=40$ m) then, $\Delta \theta_{\rm max}\approx 100$ mas ;
\item the spectral sampling of the closure phase. As already detailed by Absil et al. 2011 \cite{Absil11}, the period in the closure phase signal is roughly given by $P_{\lambda} = \lambda^2/(B\Delta\theta -\lambda)$\footnote{This formula can be obtained if we consider that the periodicity in the square visibility and in the closure phase is the same and that we have a mean baseline $B$. Then, we determine $\Delta\lambda$ using the $2\pi$ periodicity of $\sin(2\pi B\theta/\lambda)$, i.e. $2\pi B\theta/\lambda= 2\pi +2\pi B\theta /(\lambda +\Delta \lambda)$.} and must be larger than four times the spectral channel size, here equal to 0.1$\mu$m, for a proper sampling. This lead to a $\Delta\theta_{\rm max}$ of 140 mas for the mean baseline of $B=20$ m. If we consider the maximum baseline ($B= 40$ m) then we have $\Delta\theta_{\rm max}\approx85$ mas. We will thus consider a maximum field of view of about 100 mas.
\end{enumerate}

\begin{figure}[!t]
\begin{center}
\includegraphics[scale=0.6]{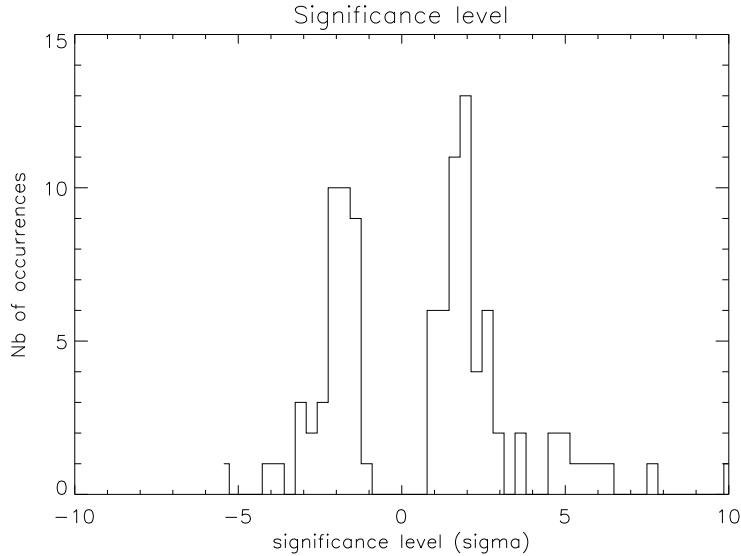}
\caption{Statistics of the significance level for the 92 stars based on the closure phases and square visibility combined. We have 4 stars with a significance level lower than 3$\sigma$: HD142, HD1581, HD56537, HD210418 which will be discuss later. In the positive part of the histogram, we have 13 potential detections which will also be discuss later (including, 5 stars with a significance level of the detection higher than 10$\sigma$ that are not represented here for clarity)}
\label{histosignt3}
\end{center}
\end{figure}

Once we have compared the data with the model, we try to determine whether or not a companion is detected by computing the signal-to-noise ratio (referred to as ``significance level'' hereafter, and expressed in terms of the estimated noise level $\sigma$), which can be associated with a given confidence level if the underlying probability distribution function is known. The first question that appears to us is which significance level to choose. A $3\sigma$ significance level was used by [\citenum{Absil11}], but here, this choice is not obvious because a significant fraction of stars (12 out of 92) has a significance level between 3 and 4$\sigma$ when we use the closure phase only. This is an additional reason why we use the combined $\chi^2$. Indeed, doing this, we can see (Fig.\ref{histosignt3}) that there are not so many stars left with a significance level higher than 3$\sigma$. One can also notice in Fig.\ref{histosignt3} that there are negative significance levels of detection which could be hard to understand. The reason is that the closure phases do not care about the sign of a companion (a negative companion creates the same closure phase as a positive companion on the other side of the star), while the effect of the companion on the squared visibilities changes sign when the contrast becomes negative. In our models, we thus consider positive or negative companions, the last case being a nonphysical one but allowing us to determine the noise distribution in our data. Actually, due to random noise, the square visibility will fluctuate around the theoretical value, sometimes lower, sometimes higher. If we add only positive companions, the squared visibility will always be lower than the visibility of a single star, and we will never manage to fit all statistical fluctuations. By adding a negative companion, we can fit the statistical fluctuations where the square visibility is higher than the theoretical one. And so, we can use this part of the histogram as a reference: we clearly see that there are still small fluctuations above 3$\sigma$ in the negative part which means that a part of the detections above 3$\sigma$ in the positive one could be due to the noise. Furthermore, this allows us to see that we have a non-Gaussian distribution of the significance level. Indeed, we can clearly see that there is no detection with a significance level of 0$\sigma$ (which would mean a single star without a companion) and that can be explained by the fact that, due to statistical fluctuations in the data, placing a companion somewhere in the field of view will always allow us to find a better fit to the data than with no companion at all. In conclusion, we take the 3$\sigma$ level as a criterion but we are aware of the fact that some of the detections could be false positives thanks to this part of the histogram.

%%-----------------------------------------------------------
\section{RESULTS}  \label{sec:results}

\begin{figure*}[!t]
\begin{center}
  \includegraphics[scale=0.30]{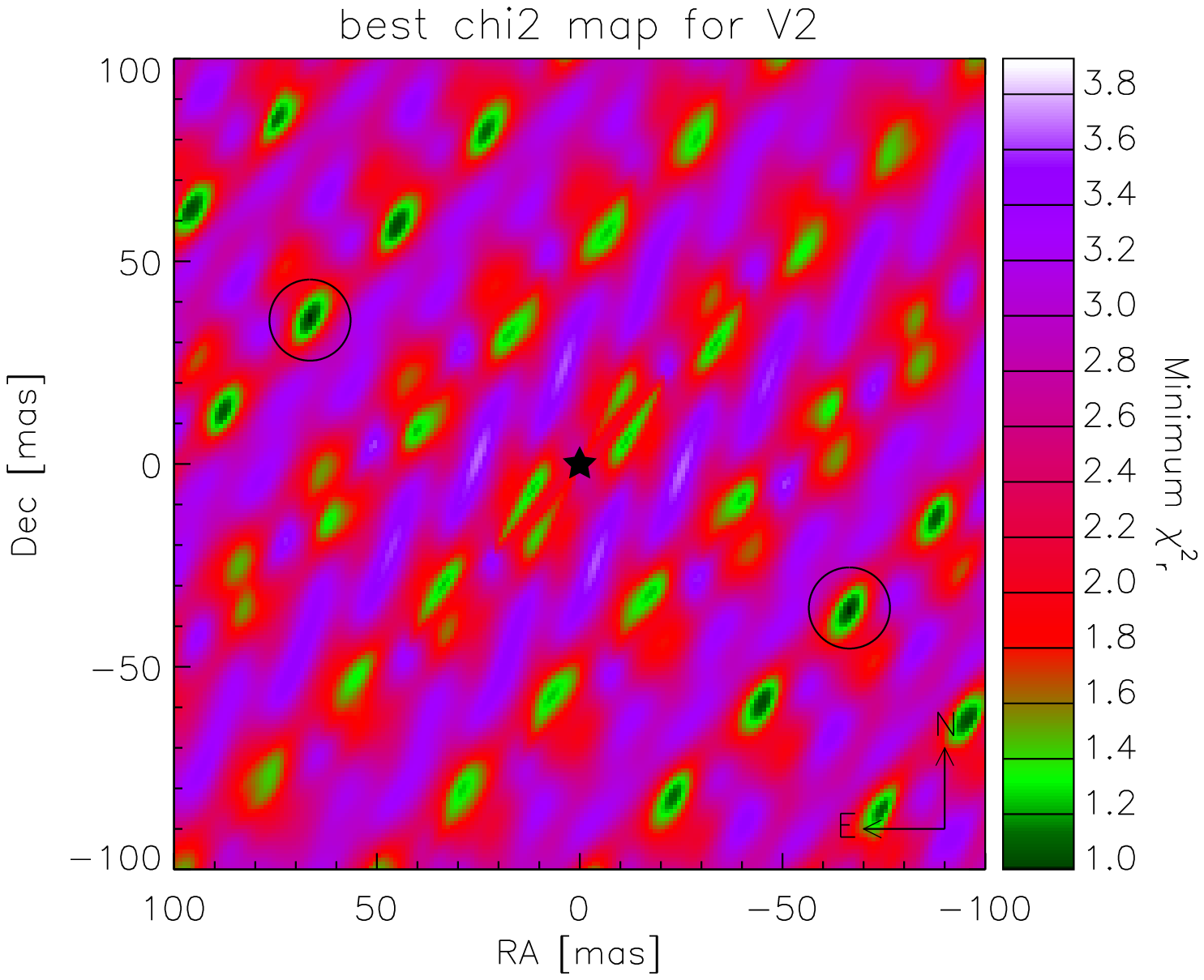}\quad
	\includegraphics[scale=0.30]{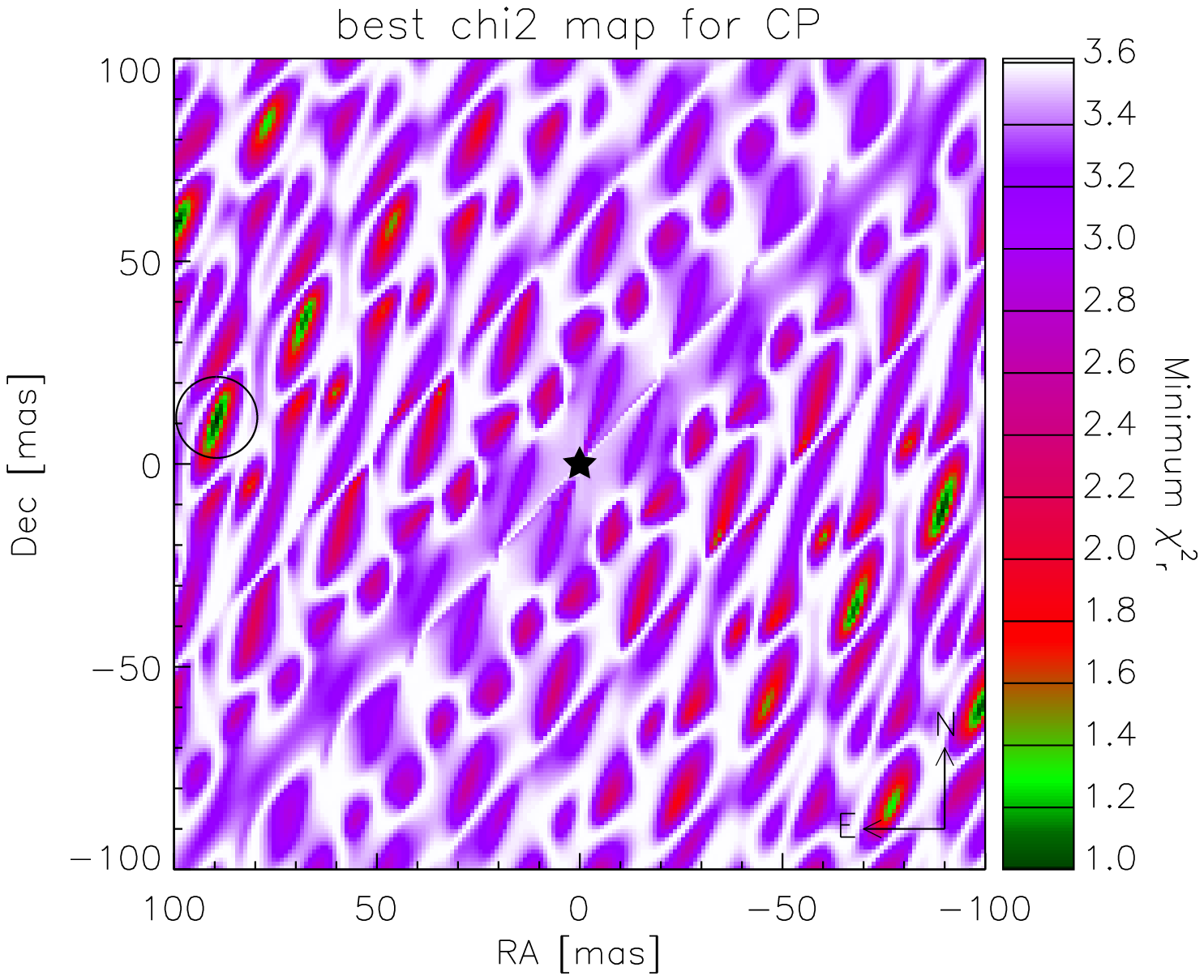}\quad
  \includegraphics[scale=0.30]{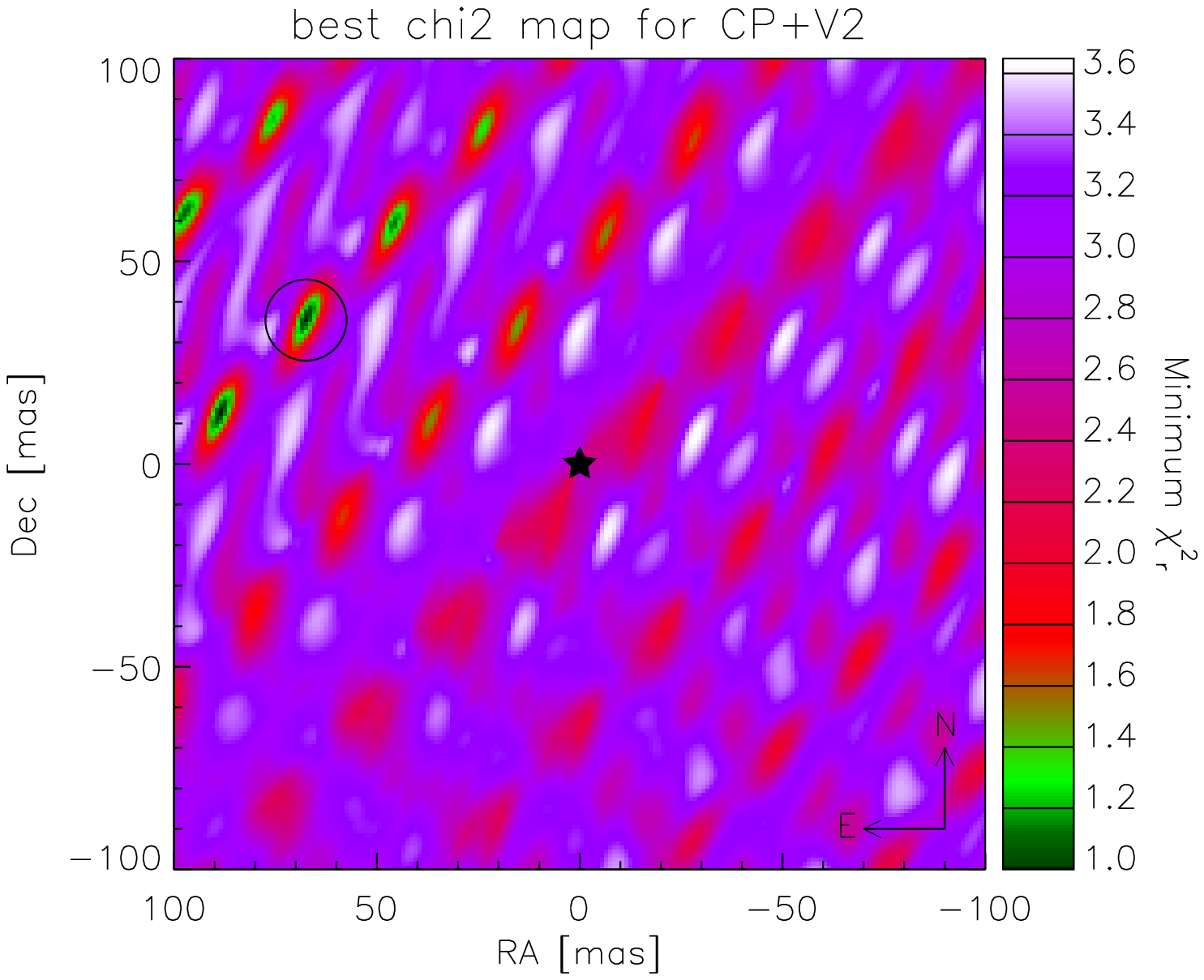}\\
	\includegraphics[scale=0.30]{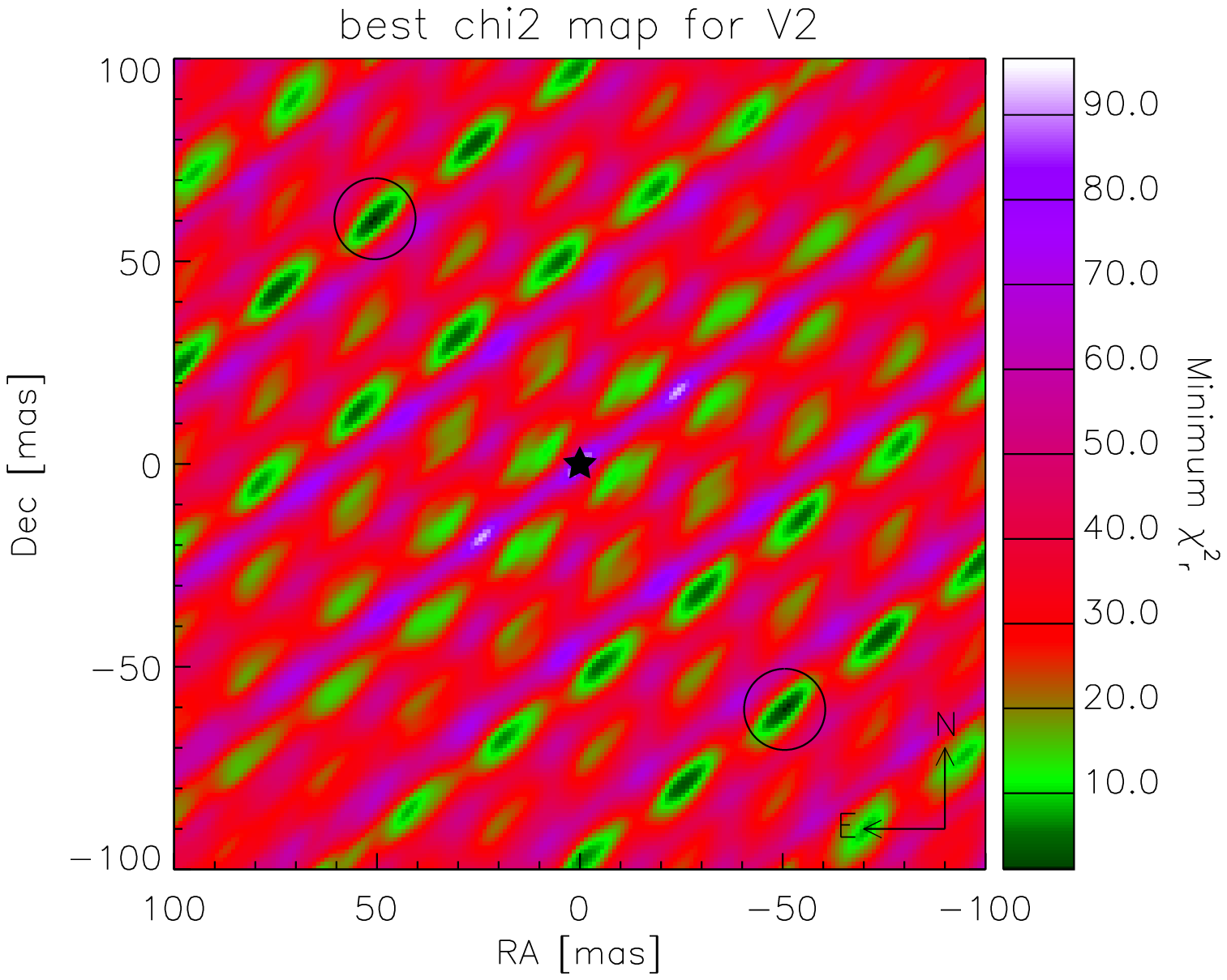}\quad
	\includegraphics[scale=0.30]{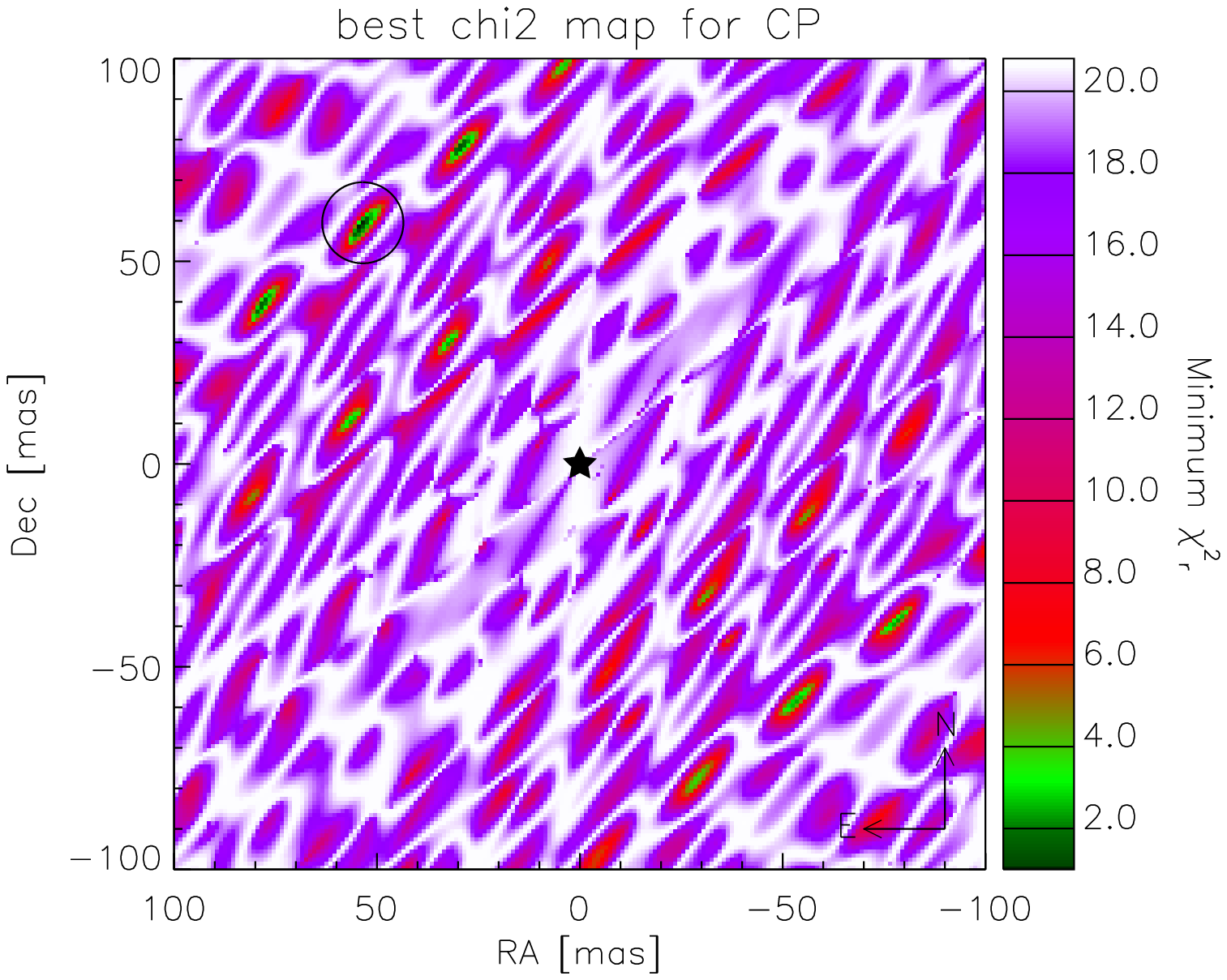}\quad
  \includegraphics[scale=0.30]{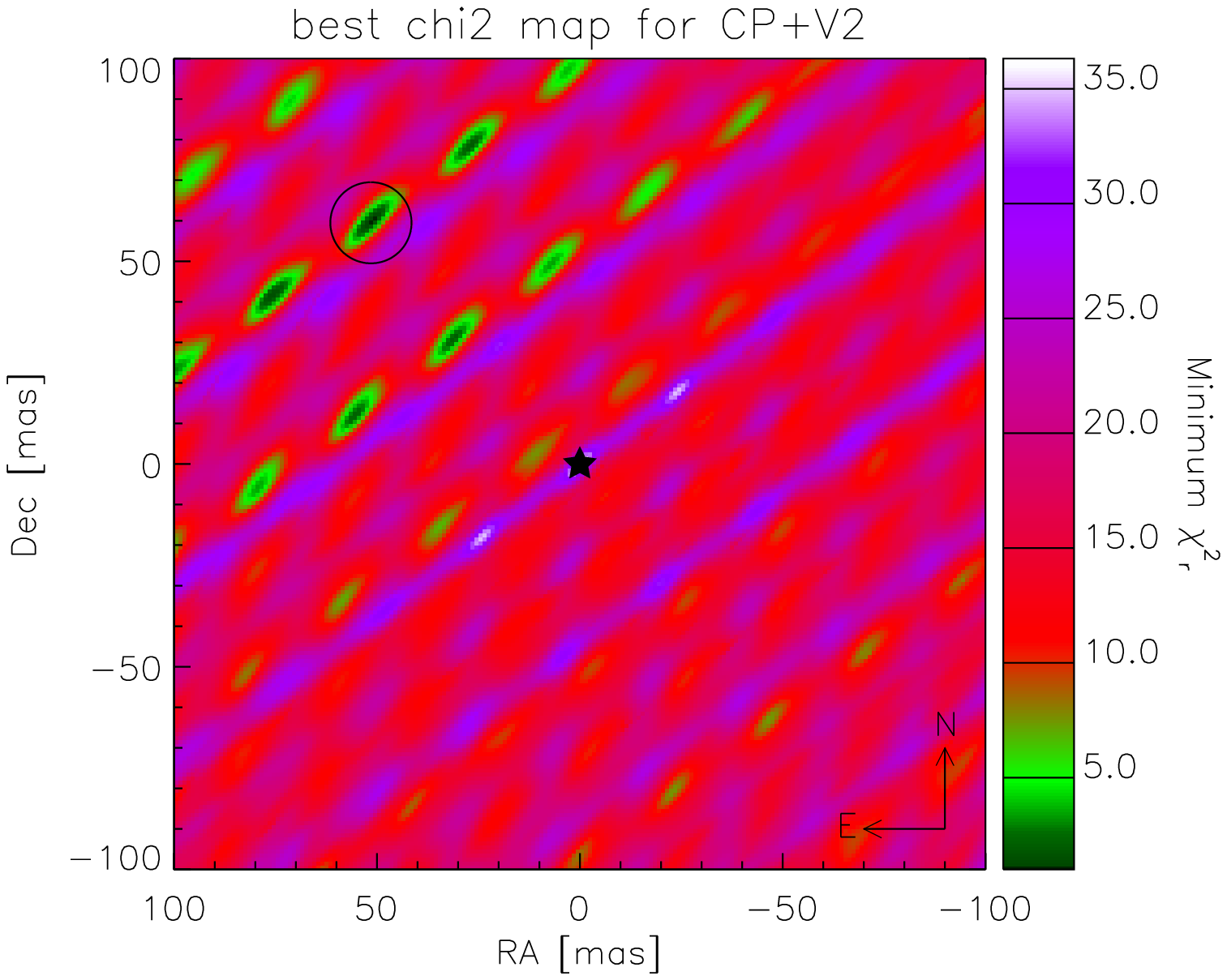}\\
	\includegraphics[scale=0.30]{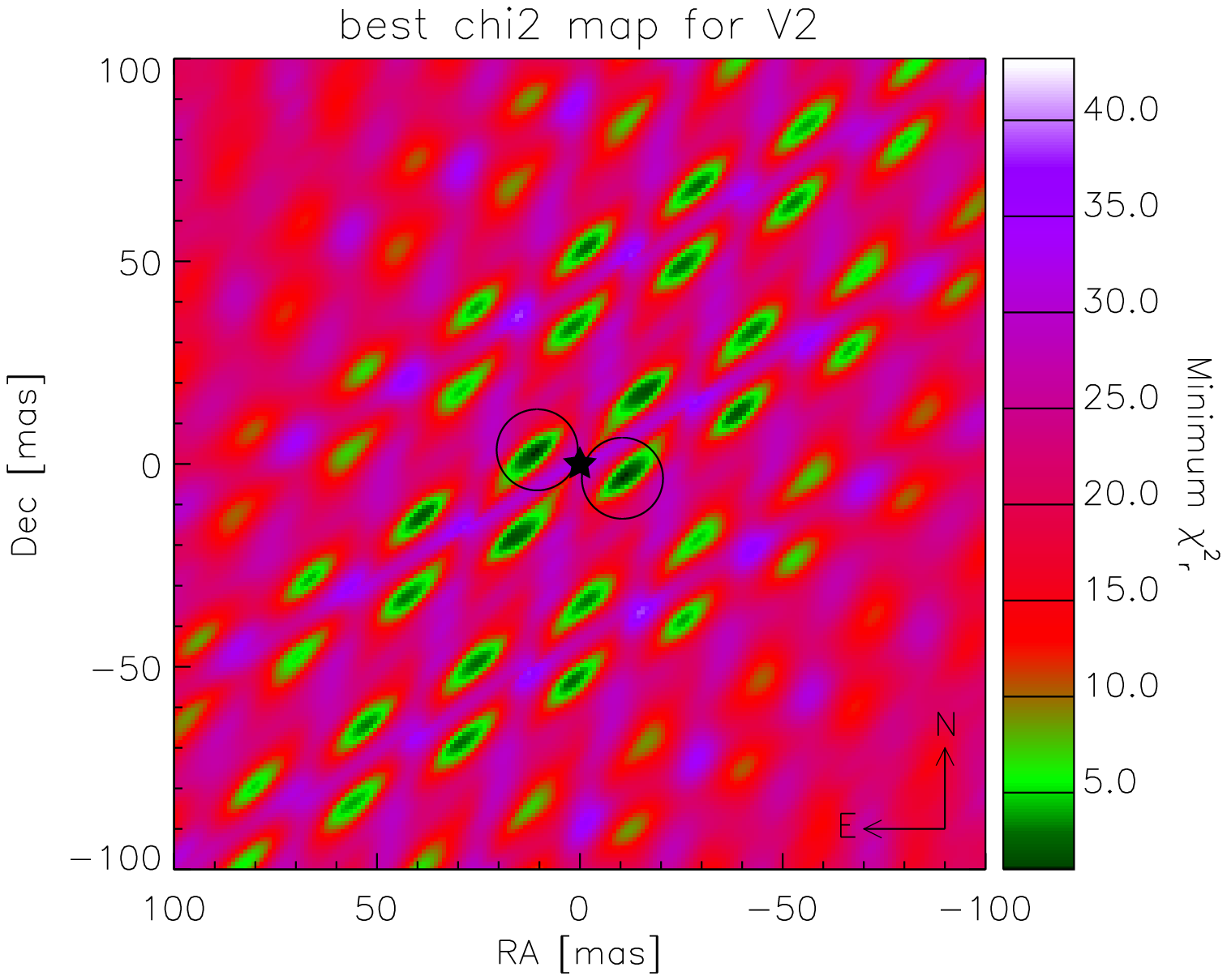}\quad
	\includegraphics[scale=0.30]{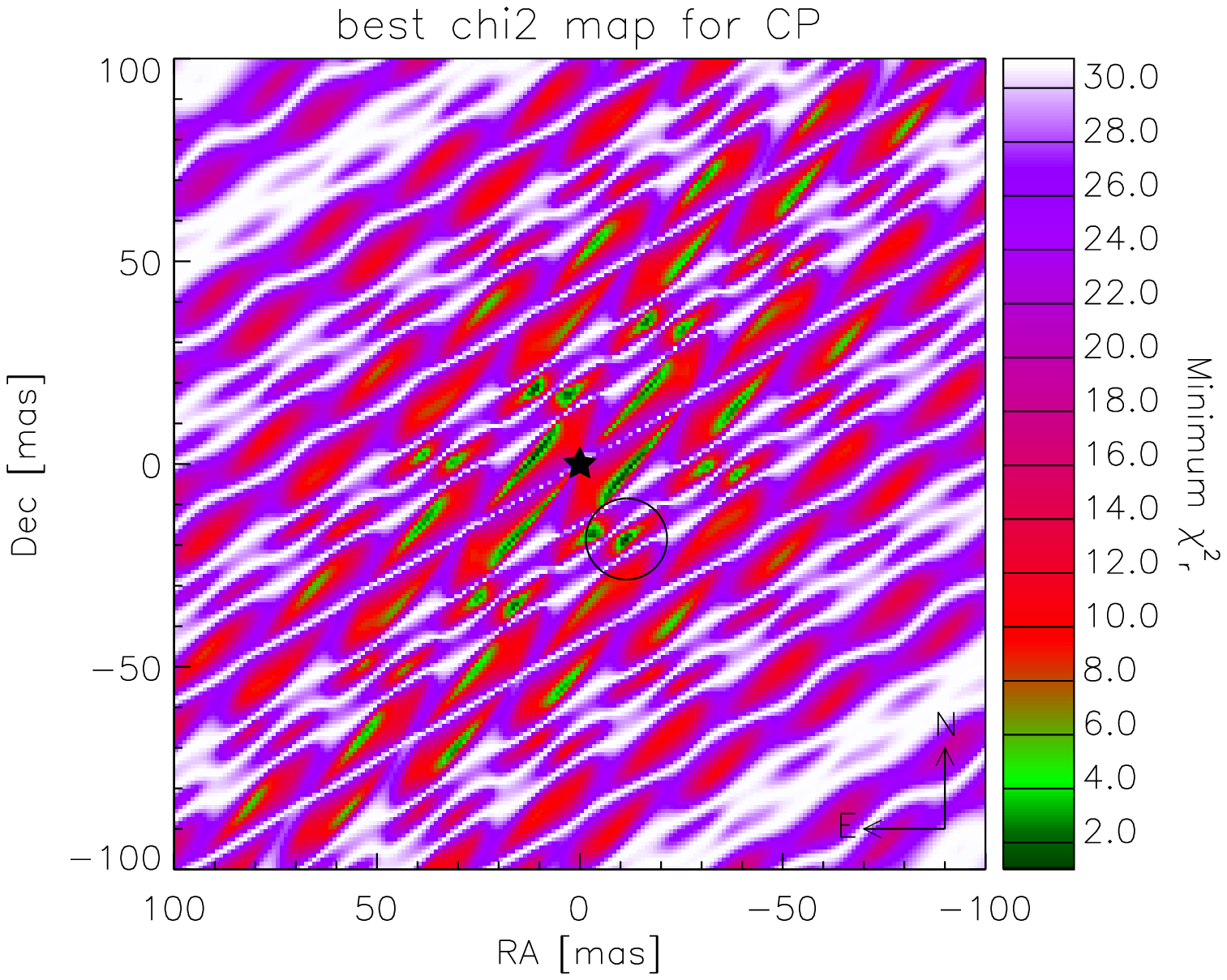}\quad
  \includegraphics[scale=0.30]{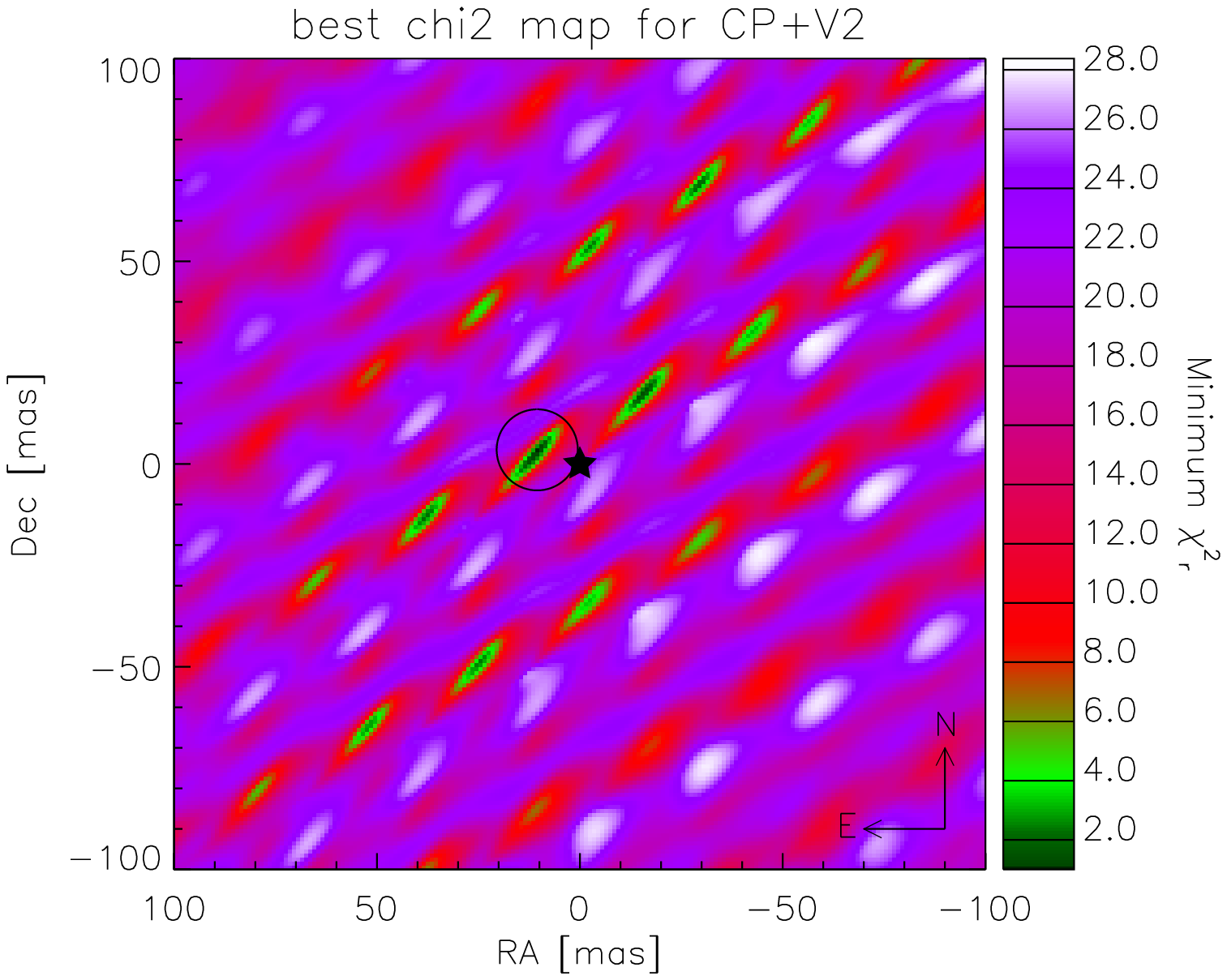}\\
	\includegraphics[scale=0.30]{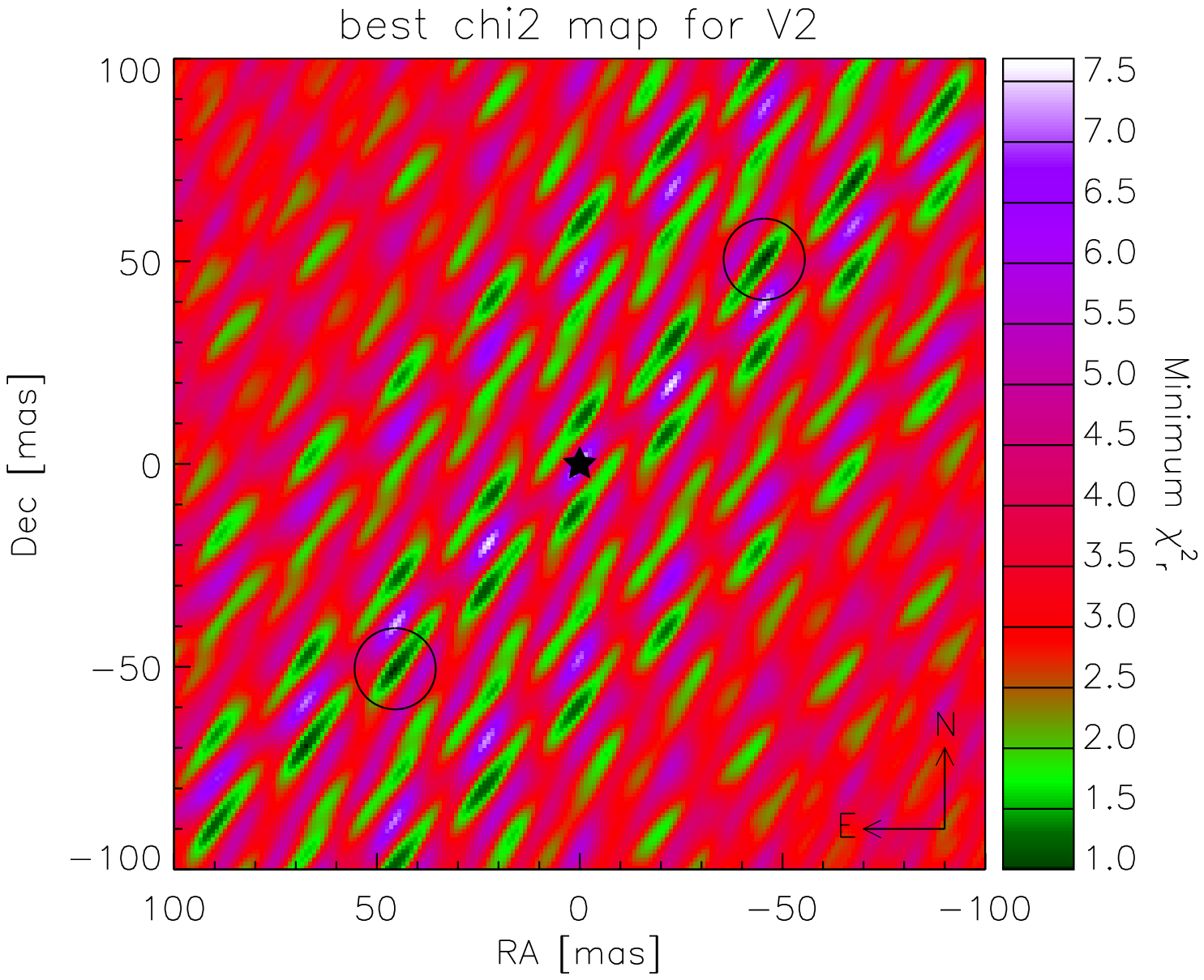}\quad
	\includegraphics[scale=0.30]{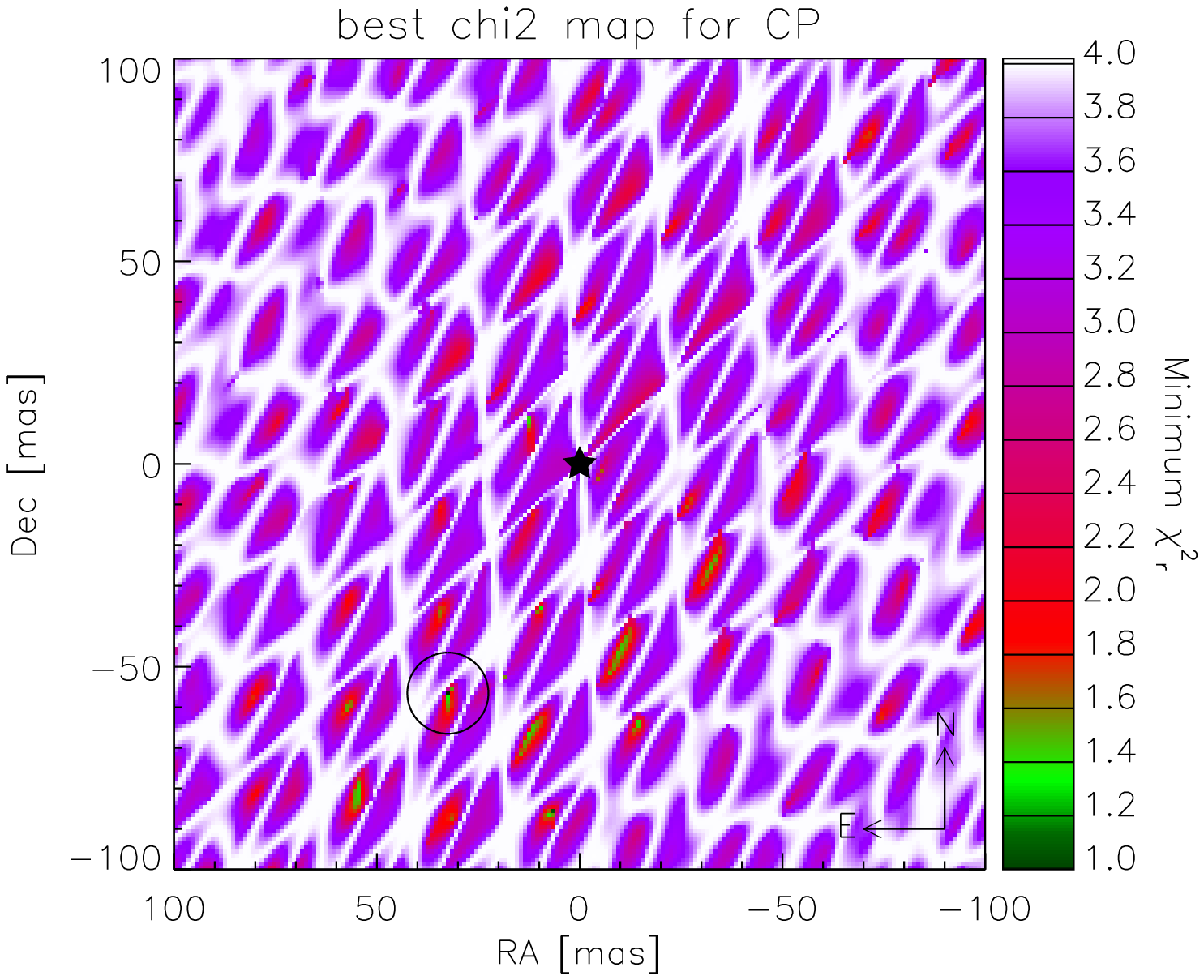}\quad
  \includegraphics[scale=0.30]{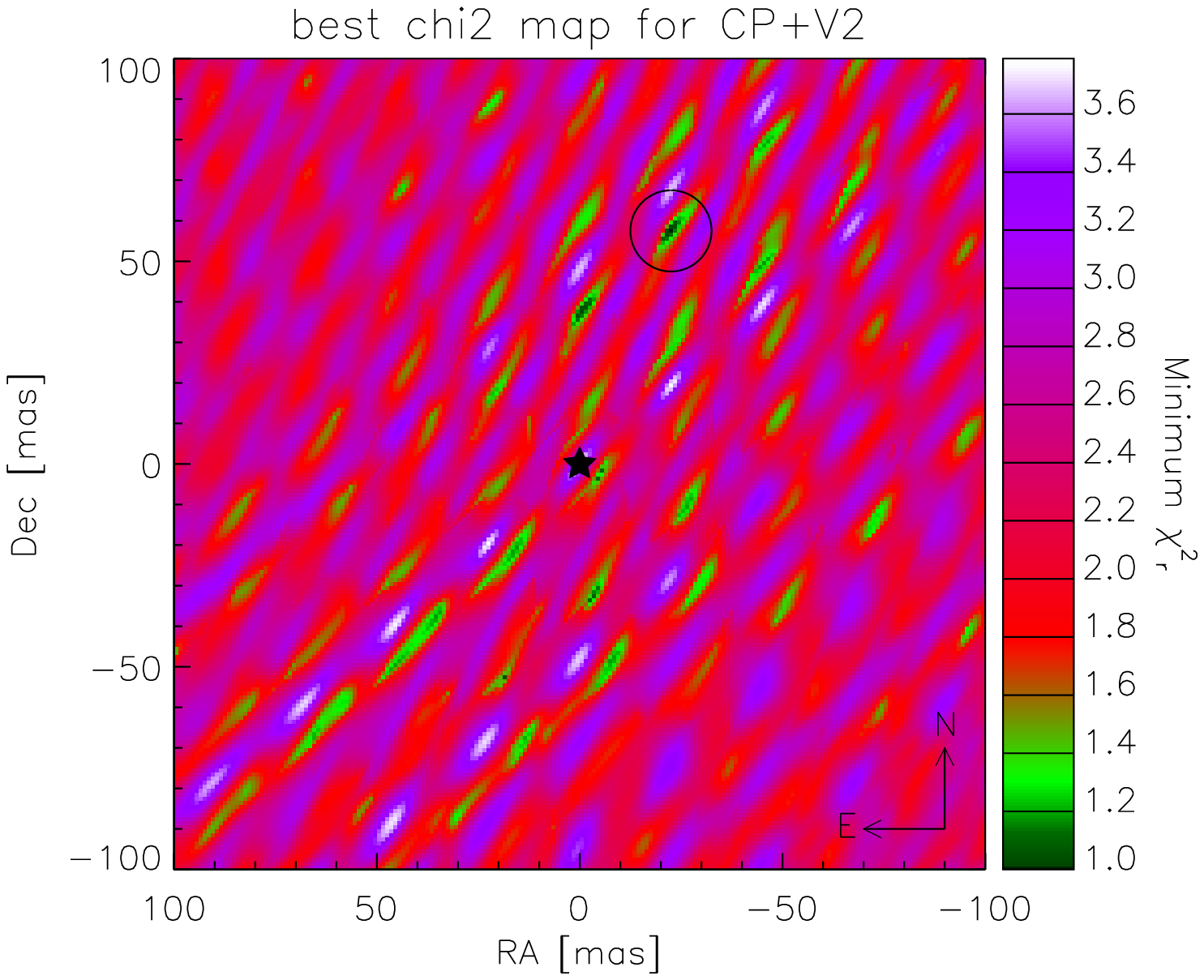}\\
\end{center}
\caption{Reduced $\chi^2$ map for the square visibility, the CP and combined square visibility and closure phase for, from top to bottom, HD4150 (date: 18-12-2012), HD16555 (date: 18-12-2012), HD29388 (16-12-2012) and HD202730 (date: 24-07-2012)}
 \label{chi2hd4150}
\end{figure*}

\begin{figure*}[!t]
\begin{center}
 \includegraphics[scale=0.30]{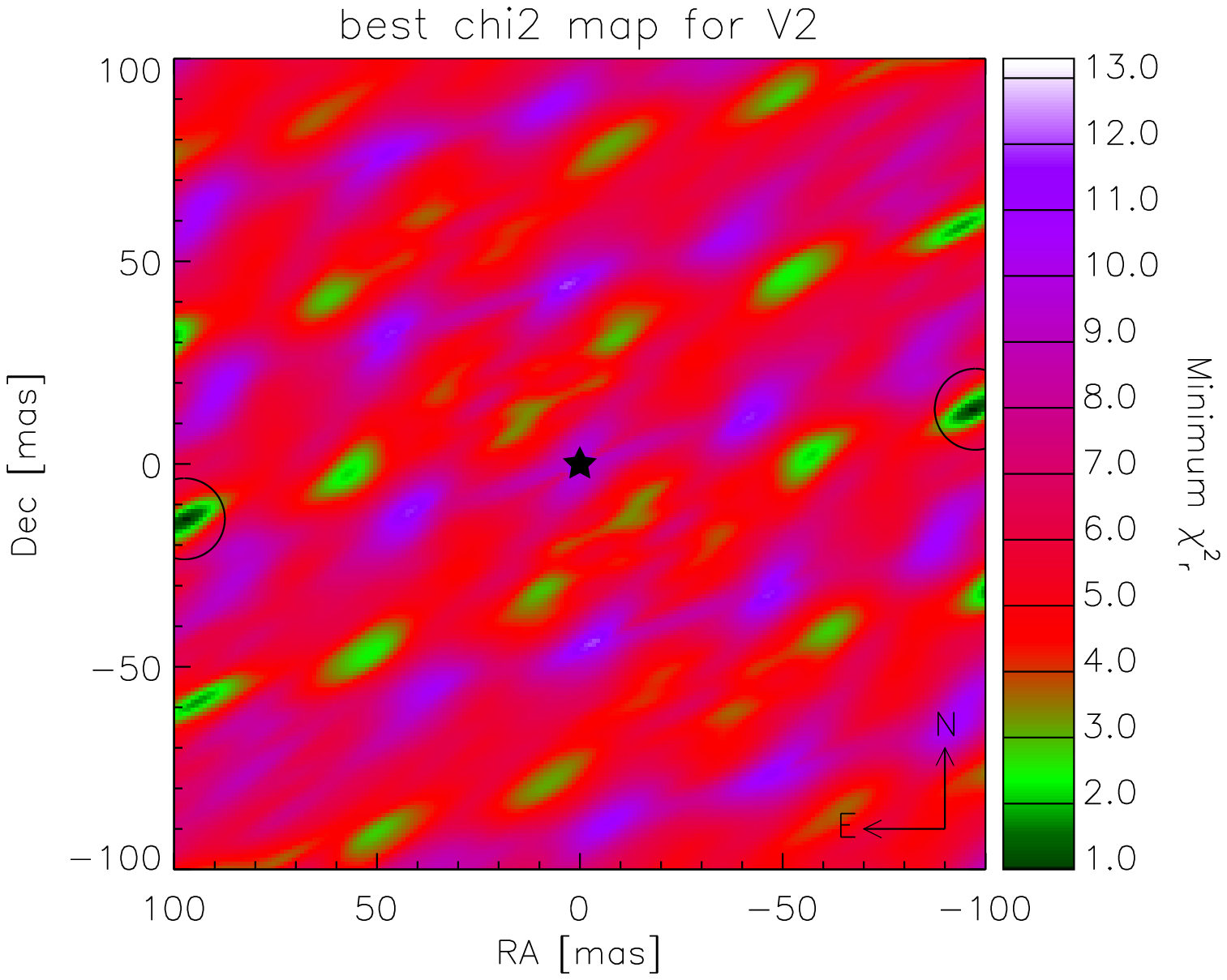}\quad
\includegraphics[scale=0.30]{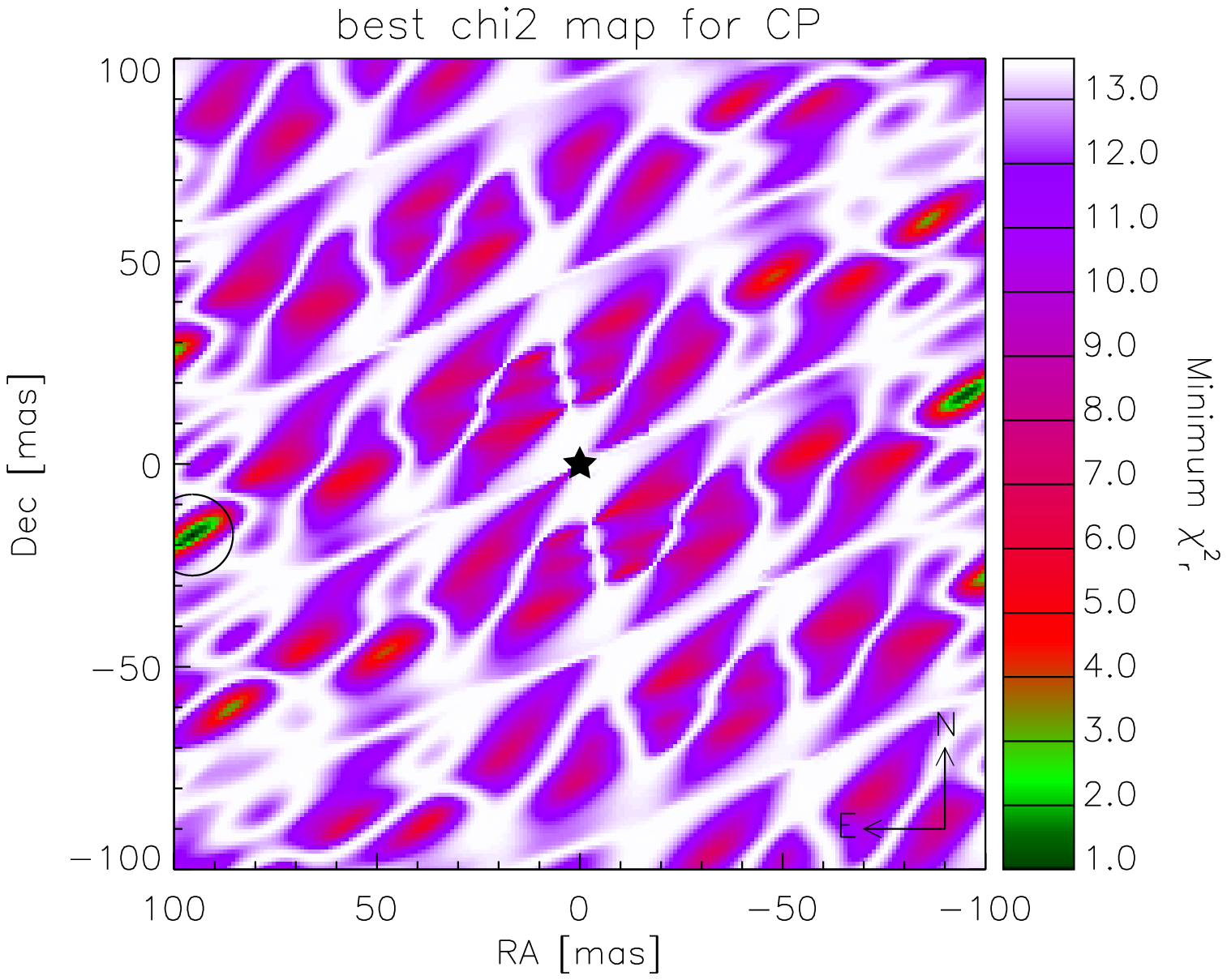}\quad
 \includegraphics[scale=0.30]{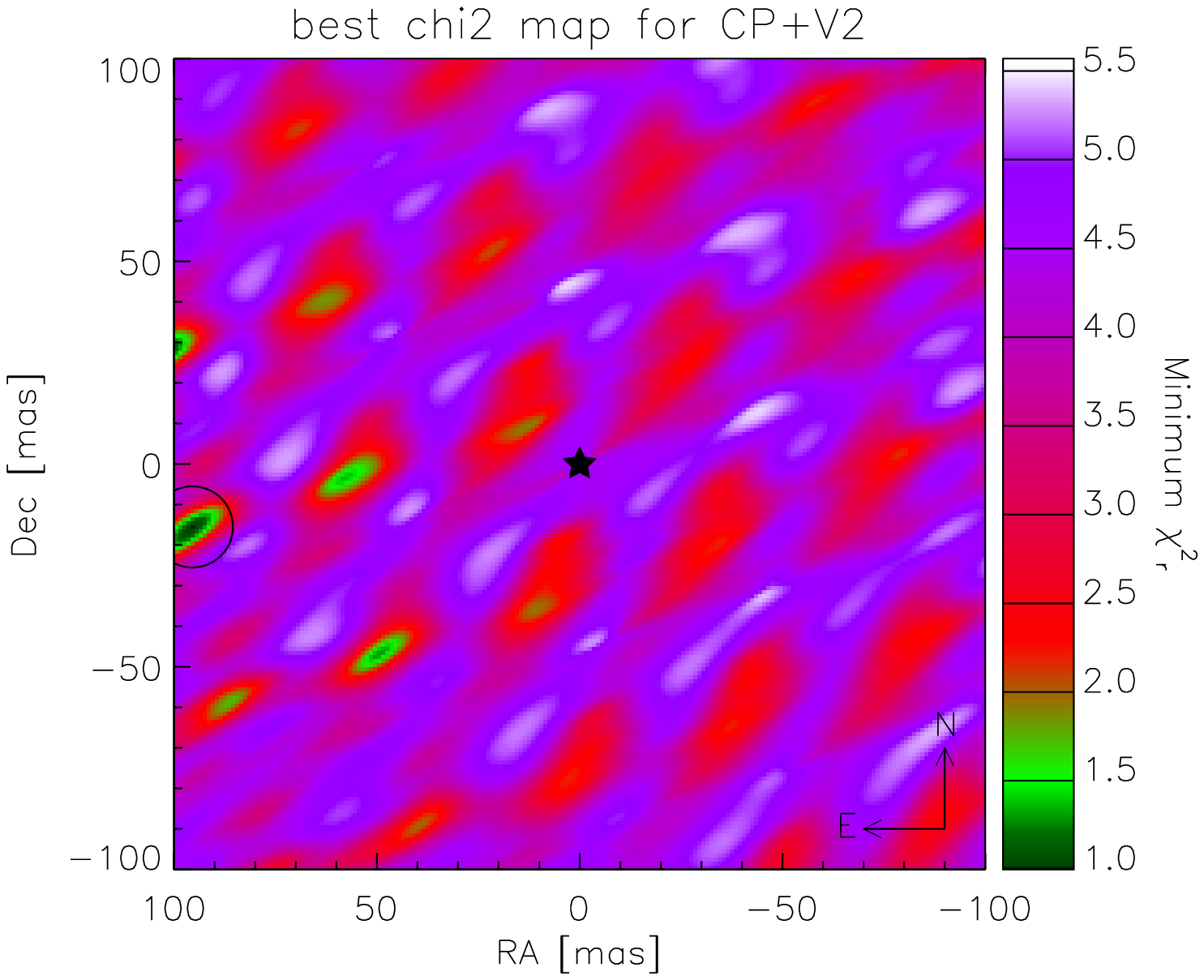}\\
\end{center}
\caption{Reduced $\chi^2$ map for the square visibility, the CP and combined square visibility and closure phase for HD4150, second epoch (08-08-2013) in K band}
 \label{fig:fitt3v2hd4150}
\end{figure*}

In order to determine the number of positive detections, we proceed as follows: we first consider all the stars with a combined significance level of detection higher than 3$\sigma$. This gives us 13 stars. Then, among those stars, we look at the closure phases to see if we have a significant signal (above $3\sigma$). This lead to the detection of a companion around five stars. The eight stars left will be considered to have an infrared excess caused by a centro-symmetric disk that only creates a signature in the square visibilities (and if this signature is really strong, this will lead to a signature in the combined chi square too). Among the five detections, four are considered as high confidence detections since the significance level is very high and the best fit position matches in the three $\chi^2$ maps. Those four detections are listed in Table \ref{tbl:detections} and the associated $\chi^2$ maps are given in the Fig.\ref{chi2hd4150}. Furthermore, a second epoch for HD4150 is also given in Fig.\ref{fig:fitt3v2hd4150} as this star has also been observed in 2013.

\begin{table*}[!t]
\caption{Summary of the detections}
\label{tbl:detections}
\begin{center}
\begin{tabular}{ccccccccc}
\hline
& & & & & & \textbf{Best fit} & \textbf{Best fit} & \textbf{Best fit} \\
\textbf{Name}  & \textbf{Date}   & \textbf{Spectral}  & \textbf{Obs.} & \textbf{Mag.} & \textbf{Distance}  &\textbf{separation} & \textbf{P.A.} & \textbf{contrast} \\
& & \textbf{type} & \textbf{band} &  & \textbf{(pc)}   & \textbf{(mas)} & \textbf{(deg)} & \textbf{(\%)} \\
\hline
HD4150 & 17-12-2012 & A0IV & H & 4.4  & 75.5 & $90.5 \pm 2.2$ & $-84.0\pm 2.2$ & $2.3 \pm 0.4$\\
$\cdots$ & 09-08-2013 & $\cdots$ & K & 4.3 & $\cdots$ & $96.8 \pm 2.4$ & $-99.2 \pm 1.1$ & $4.1 \pm 0.4$ \\
HD16555  & 18-12-2012 & A6V & H & 4.6 & 45.6 & $78.7 \pm 1.6$ & $-40.9\pm 0.3$ & $51 \pm 4$ \\
HD29388  &16-12-2012& A6V & H & 4.1 &47.1 & $11.1 \pm 0.2$ & $-71.6 \pm 0.05$ & $3.0 \pm 0.2$ \\
HD202730 &24-07-2012 & A5V & H & 4.4 & 30.3 & $61.7 \pm 1.2$ & $21.4 \pm 0.8$ & $87 \pm 14$ \\
\hline
\end{tabular}
\end{center}
\end{table*}

This leaves the fifth star, HD224392, as a special case. Indeed, it is difficult to determine if this star is effectively a binary or not based on our 2012 data, because although both the visibilities and closure phases show a significant detection, the best-fit position does not match between the two $\chi^2$ maps. This case will be further discussed, using more data sets, in a forthcoming paper (Borgniet et al., in prep).

%%-----------------------------------------------------------
\section{On the PIONIER sensitivity} \label{sec:sensitivity}

In the case of non detections (79 stars out of 92 in our sample), we can determine an upper limit on the presence of companion as a function of the position in the field-of-view, using the $\chi^2$ statistics as explained in [\citenum{Absil11}]. This map of the sensitivity level can then be used to derive the median sensitivity (i.e., sensitivity achieved for $50\%$ of the positions within the search region). We can also define the sensitivity at a higher completeness level, e.g., the sensitivity reached for $90\%$ of the positions within the search region.

\begin{figure*}[!t]
\begin{center}
 \includegraphics[scale=0.45]{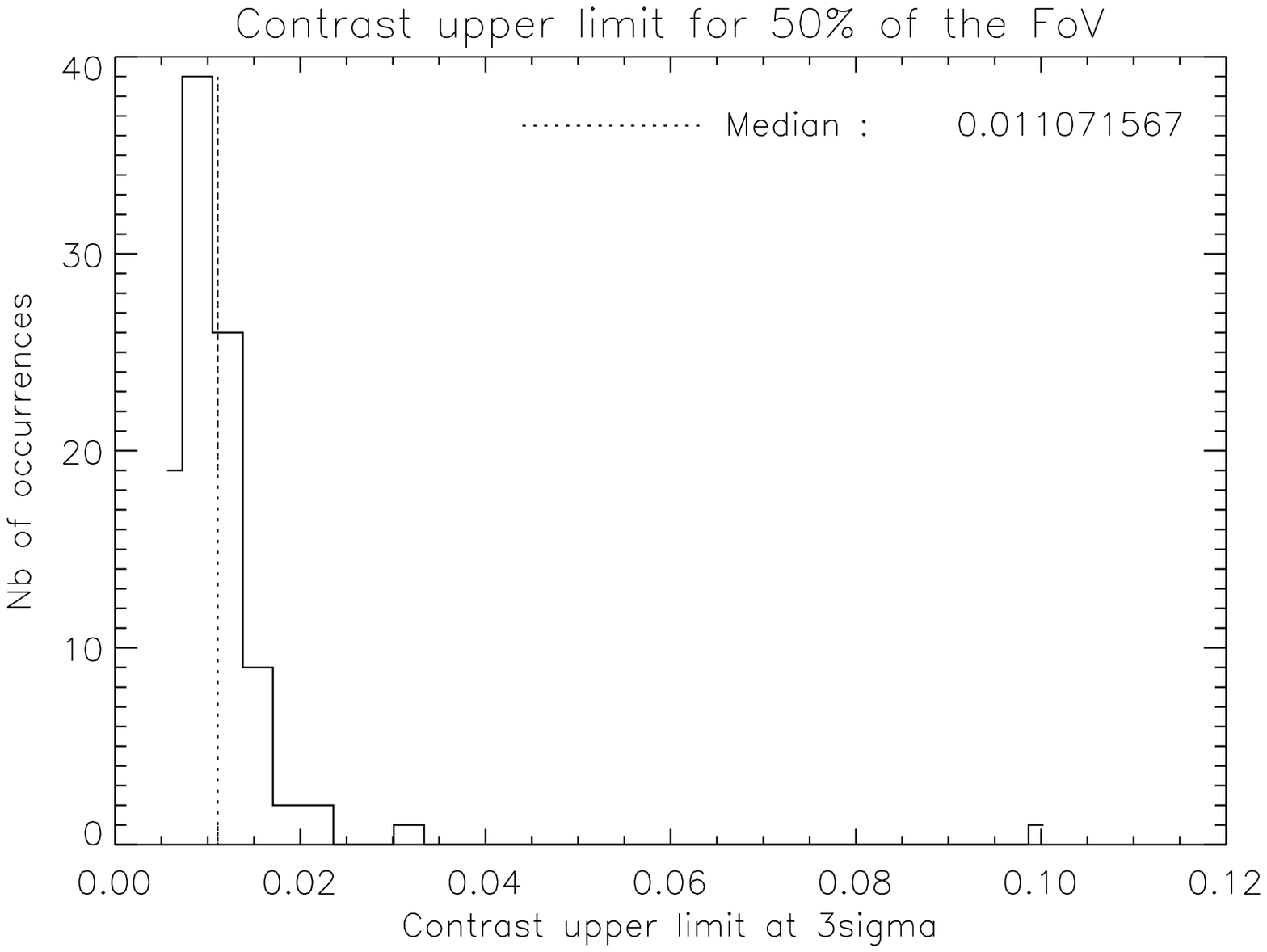}\quad
\includegraphics[scale=0.45]{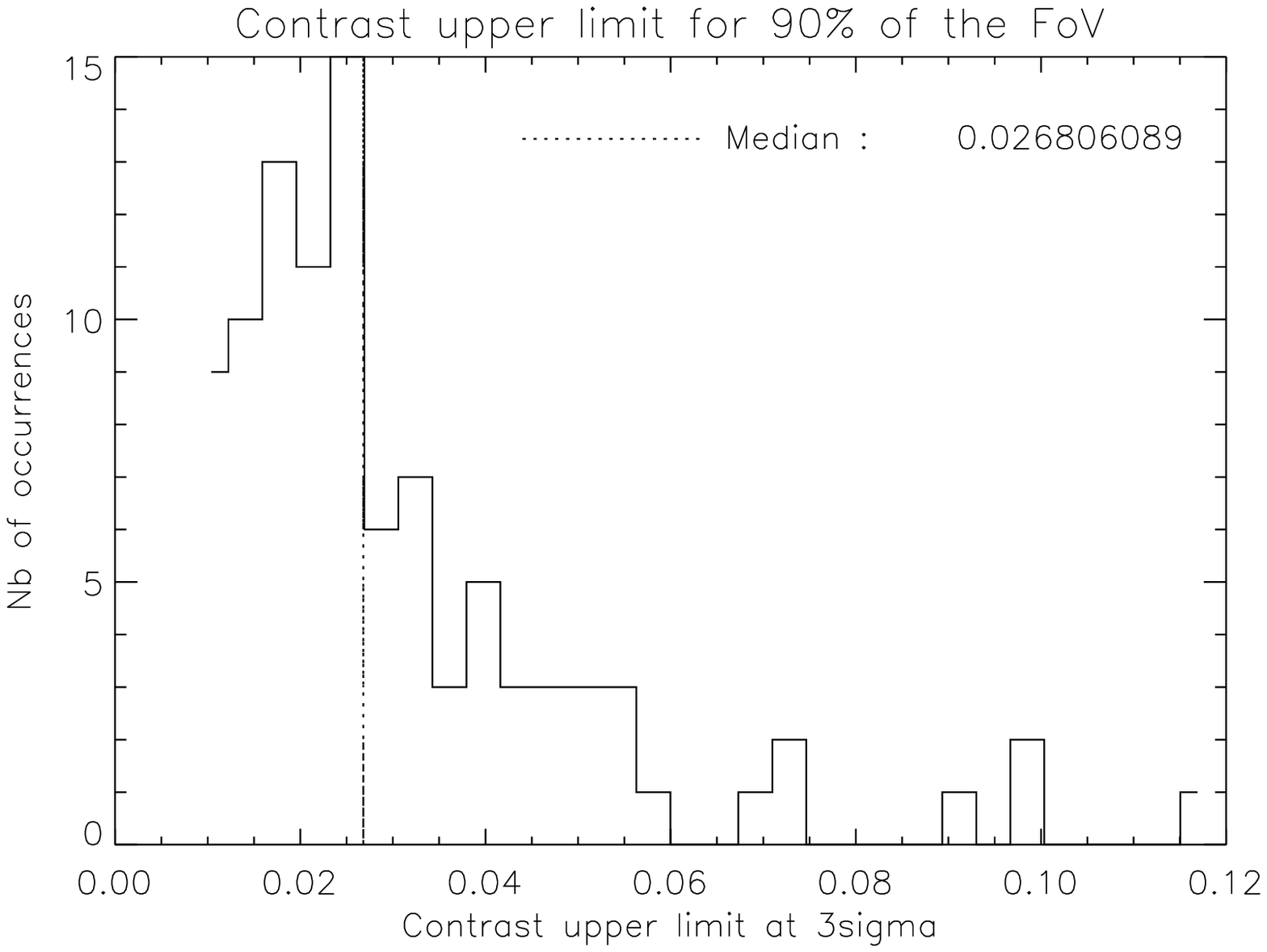}\\
\includegraphics[scale=0.45]{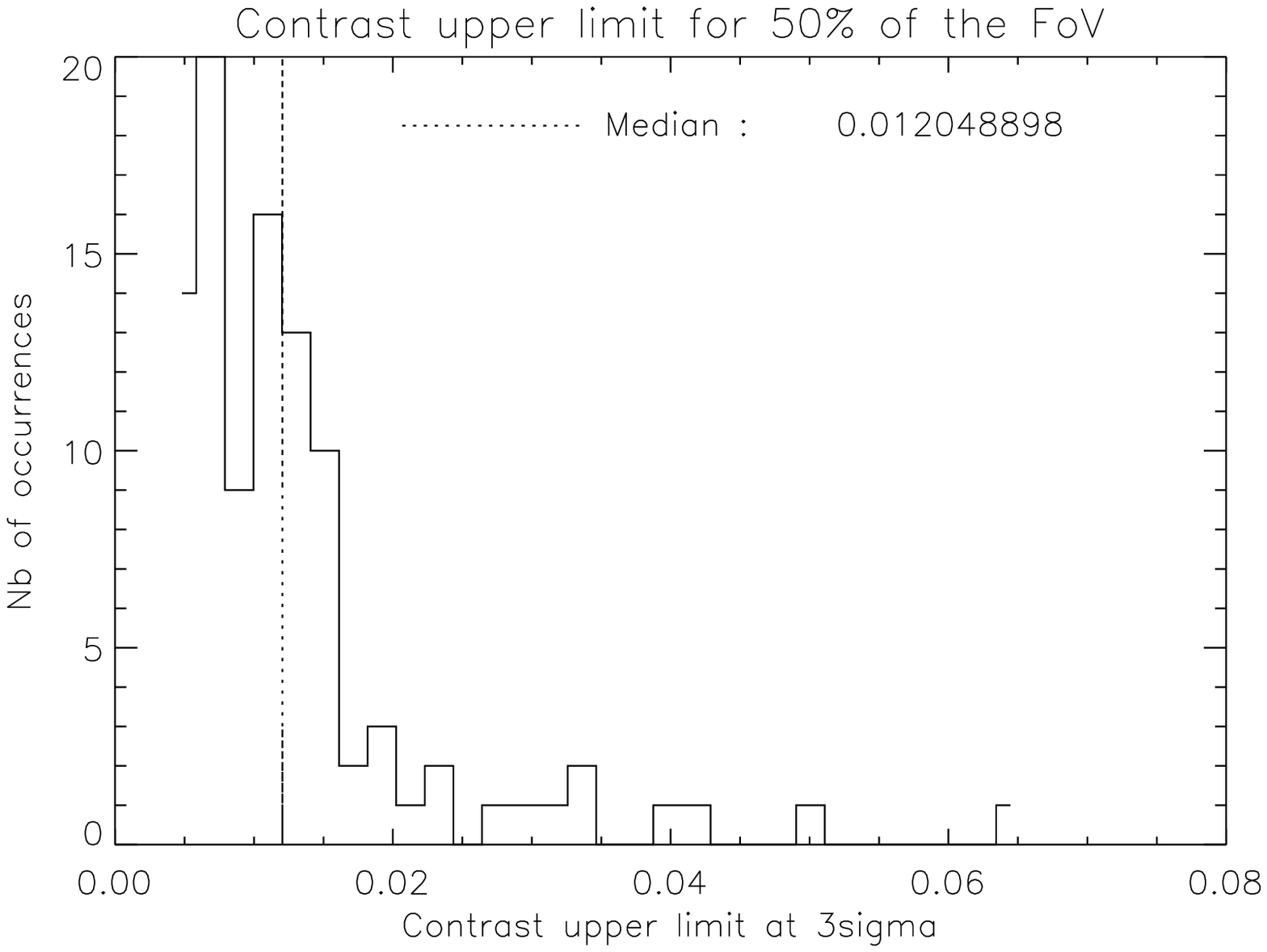}\quad
\includegraphics[scale=0.45]{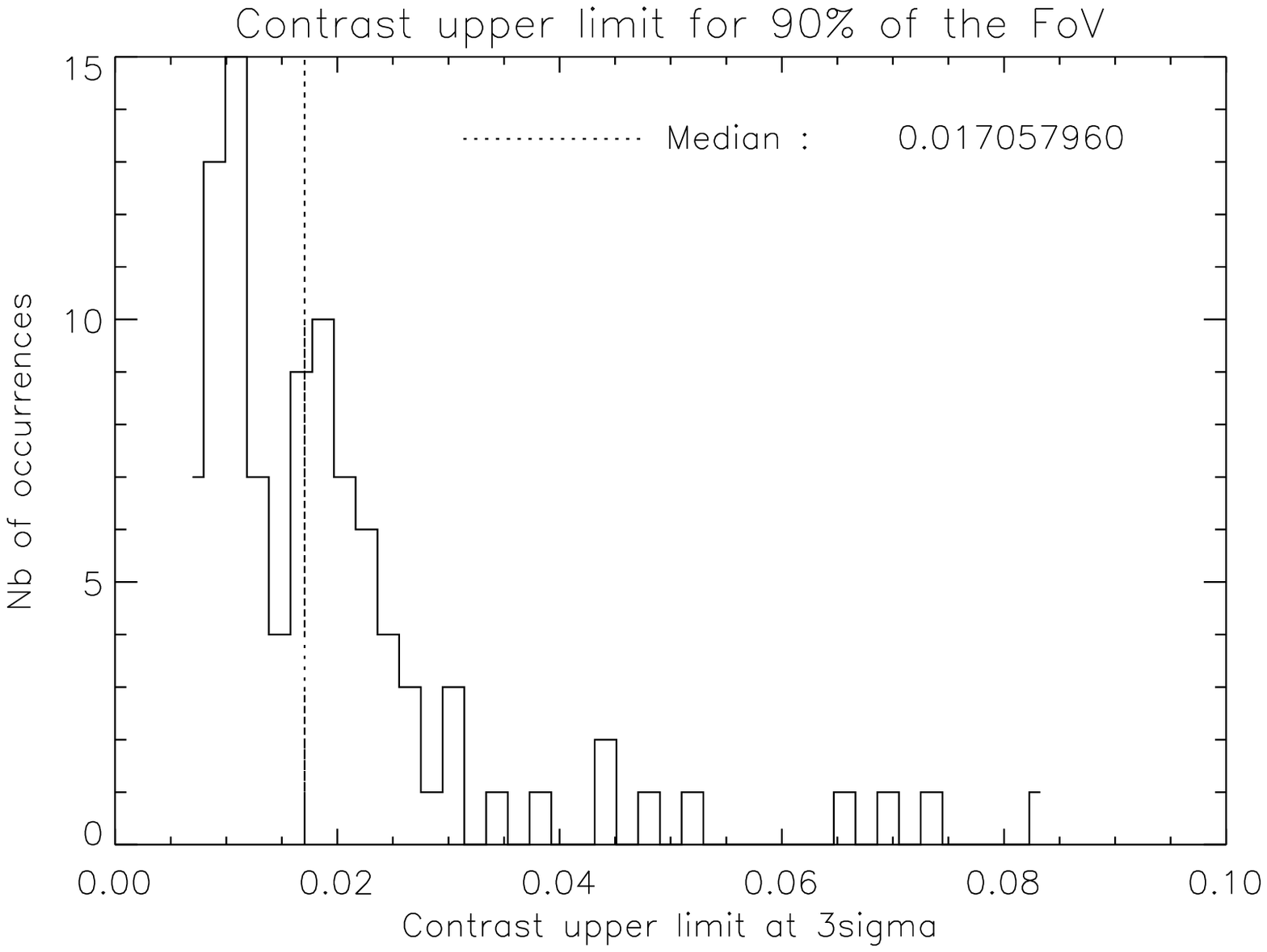}\\
\includegraphics[scale=0.45]{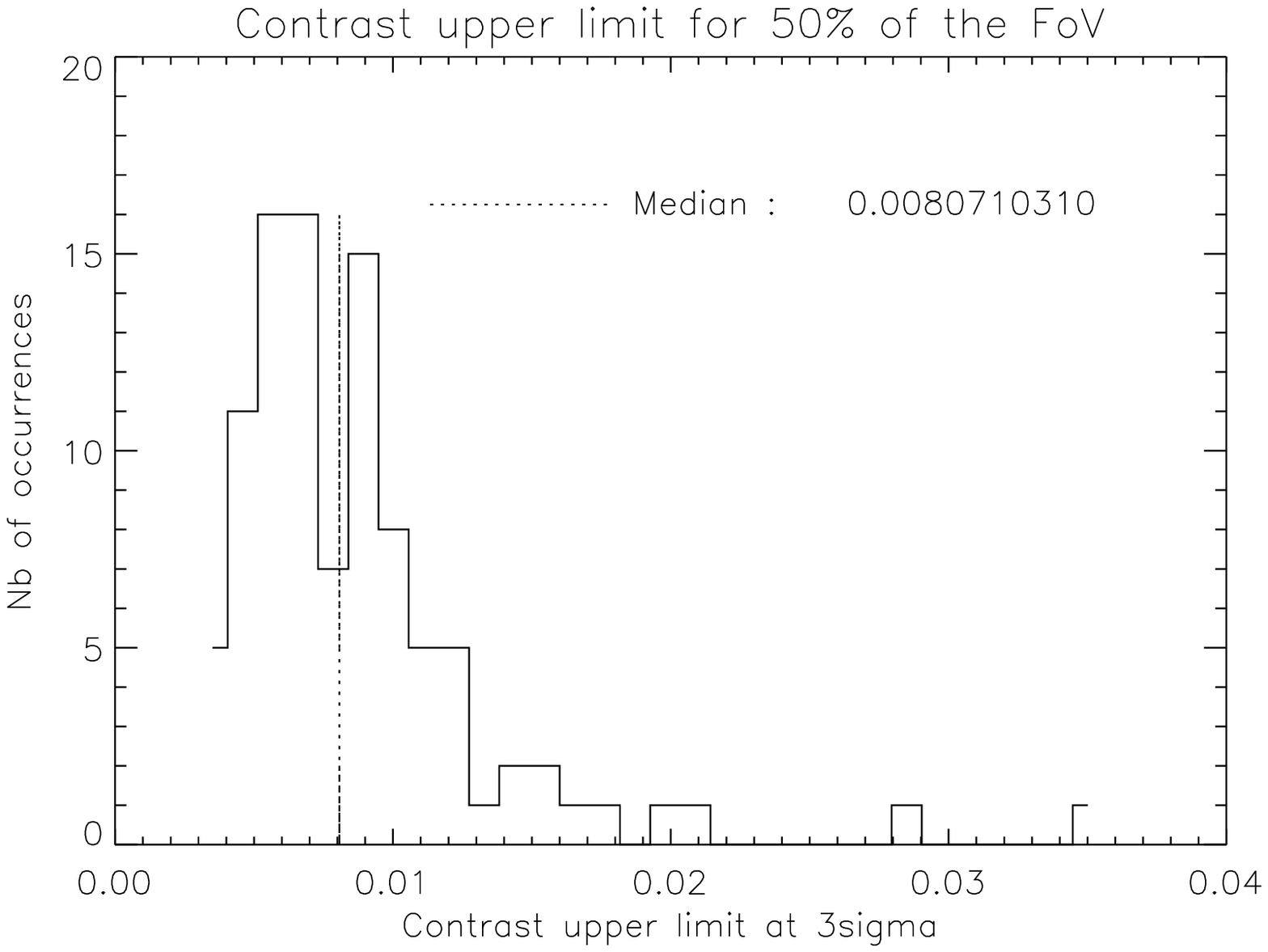}\quad
 \includegraphics[scale=0.45]{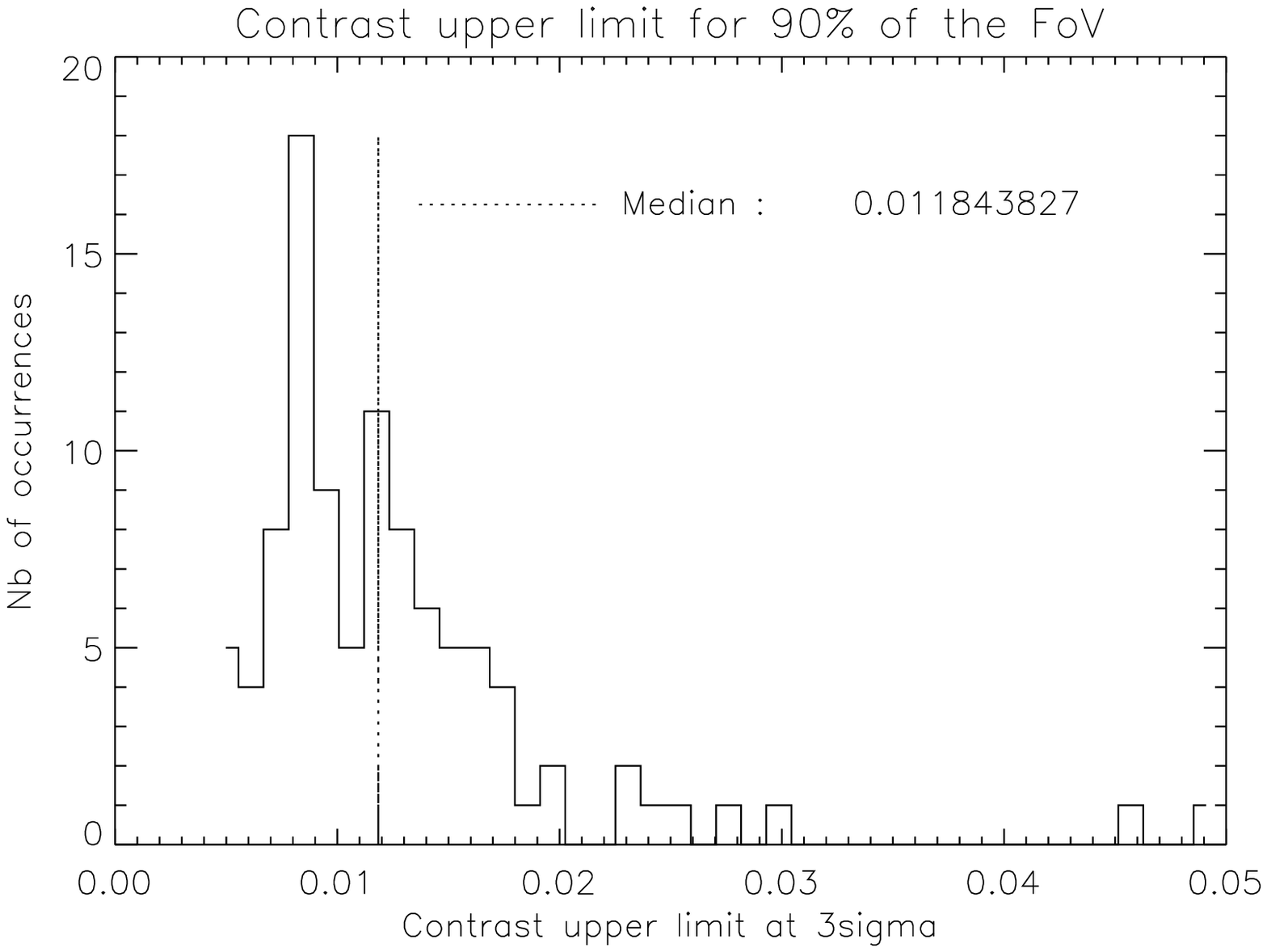}\\
\end{center}
\caption{From top to bottom: Statistic of the contrast upper limit at 3$\sigma$ for 50\%  and for 90\% of the field of view for the closure phases,
 Statistic of the contrast upper limit at 3$\sigma$ for 50\% and for 90\% of the field of view for the square visibility,
 Statistic of the contrast upper limit at 3$\sigma$ for 50\% and for 90\% of the field of view for the closure phases and the square visibility combined}
 \label{fig:sensitivitydet}
\end{figure*}

In order to deduce the typical sensitivity of PIONIER in ``survey mode'' (3 OBs per target), we produced the histograms of the sensitivity levels for all stars without a significant detection of a near-infrared excess. These histograms give us the sensitivity to companions in 50\% or 90\% of the search region in a significance level of 3$\sigma$ for the closure phases, the square visibilities and the combination of the two. They are illustrated in Fig.\ref{fig:sensitivitydet}. We deduce from this figure that  the sensitivity of PIONIER is around 1\% when using the closure phases and the squared visibilites in a combined way (median sensitivity $0.8\%$, percentile 90 sensitivity $1.2\%$).

\begin{figure*}[!t]
\begin{center}
 \includegraphics[scale=0.45]{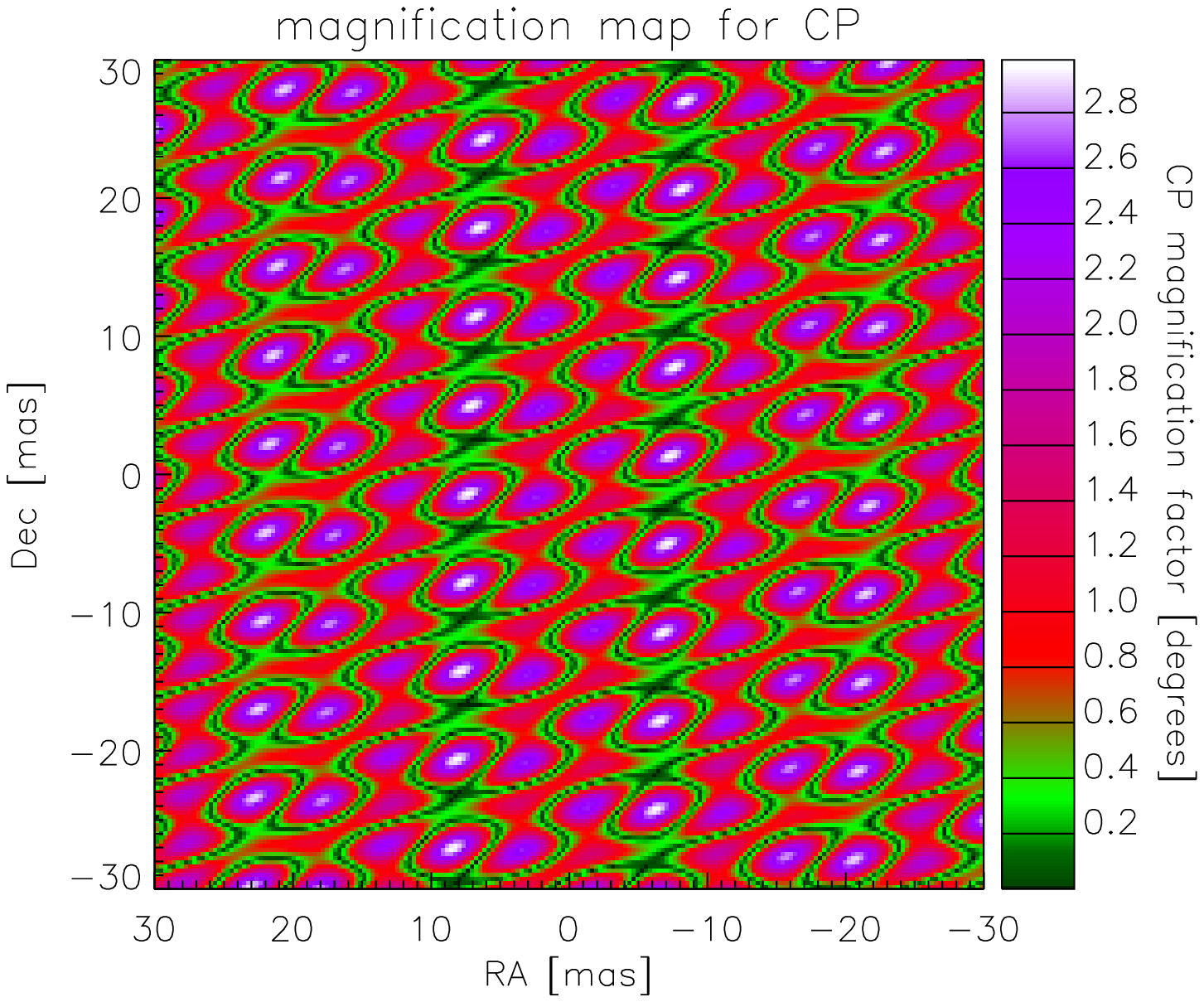}\quad
\includegraphics[scale=0.45]{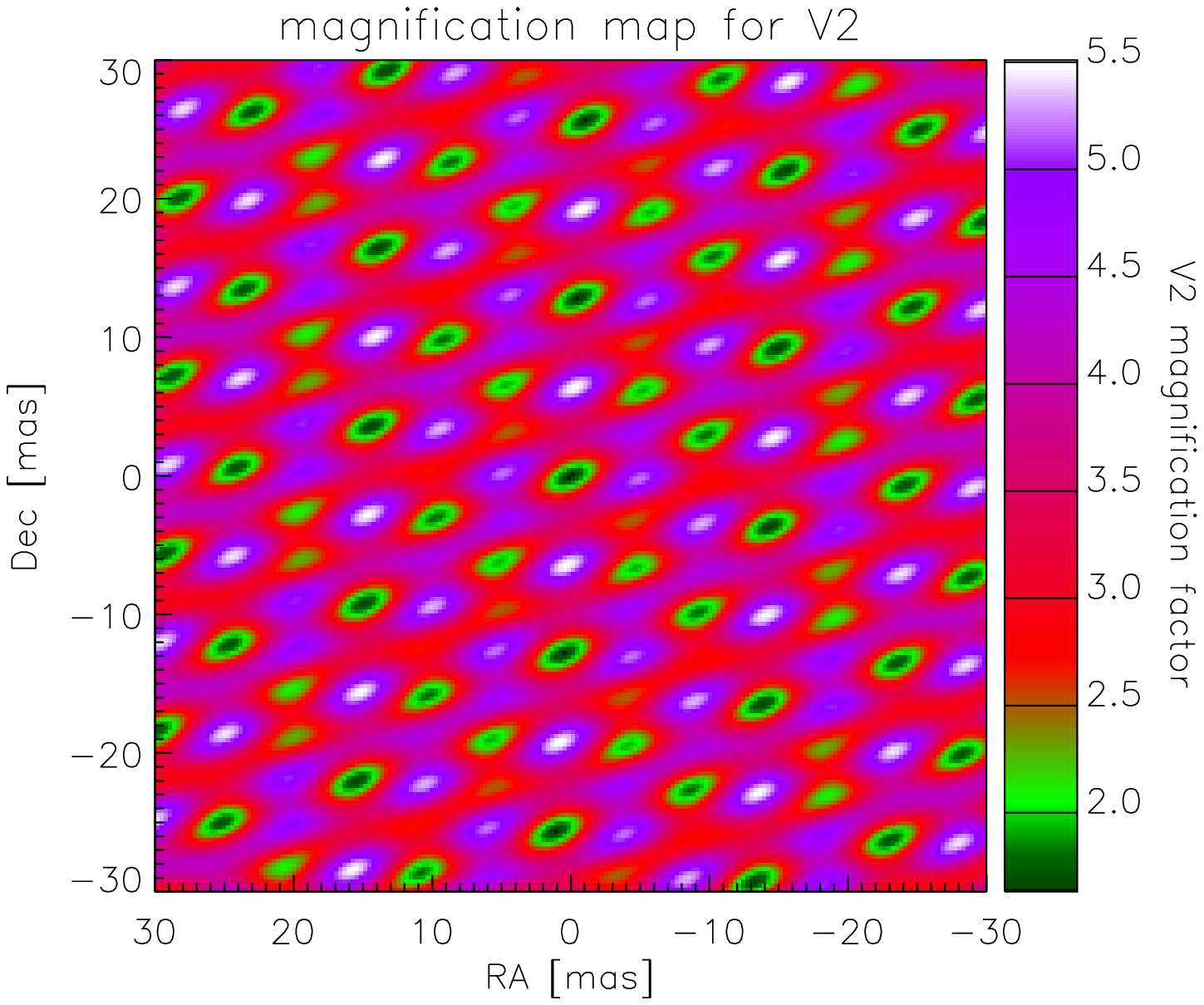}\quad
\includegraphics[scale=0.50]{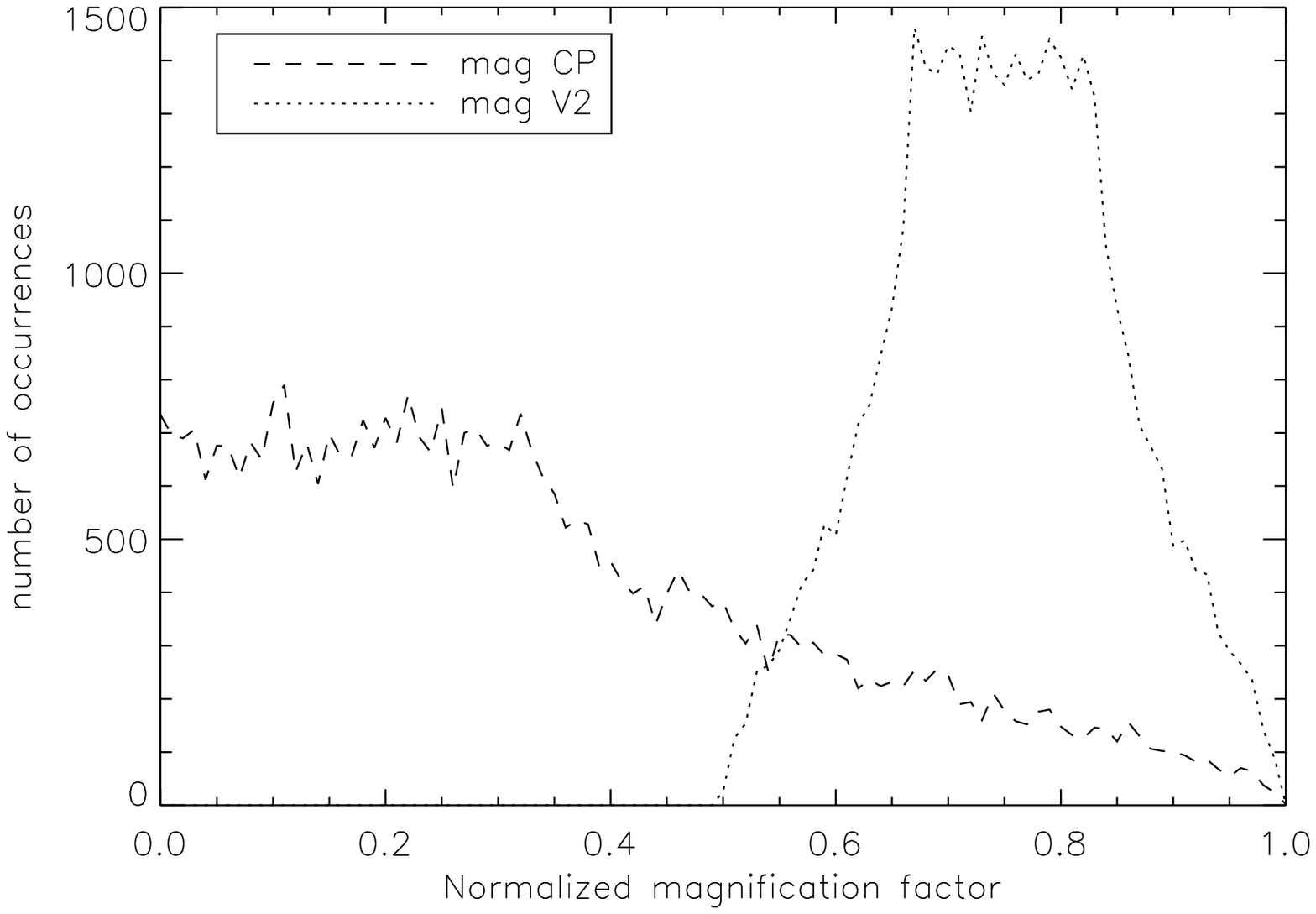}\\
\end{center}
\caption{Magnification map for the closure phase (in absolute value), the square visibility and corresponding histograms for the normalized magnification for both the closure phase and the square visibility, using a triangular array (VLTI stations A1-K0-G1).}
 \label{fig:magmaps}
\end{figure*}

\begin{figure*}[!t]
\begin{center}
 \includegraphics[scale=0.45]{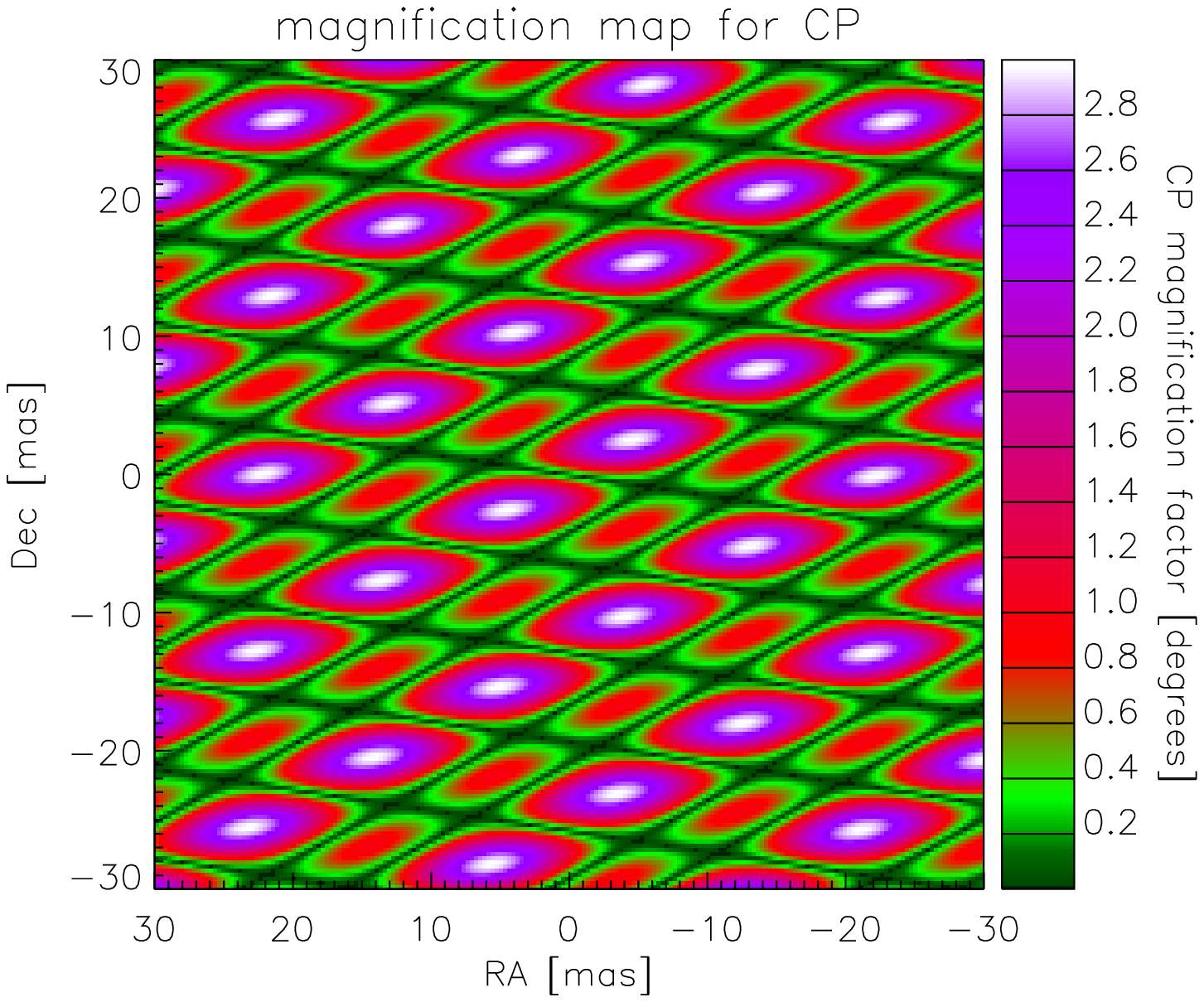}\quad
\includegraphics[scale=0.45]{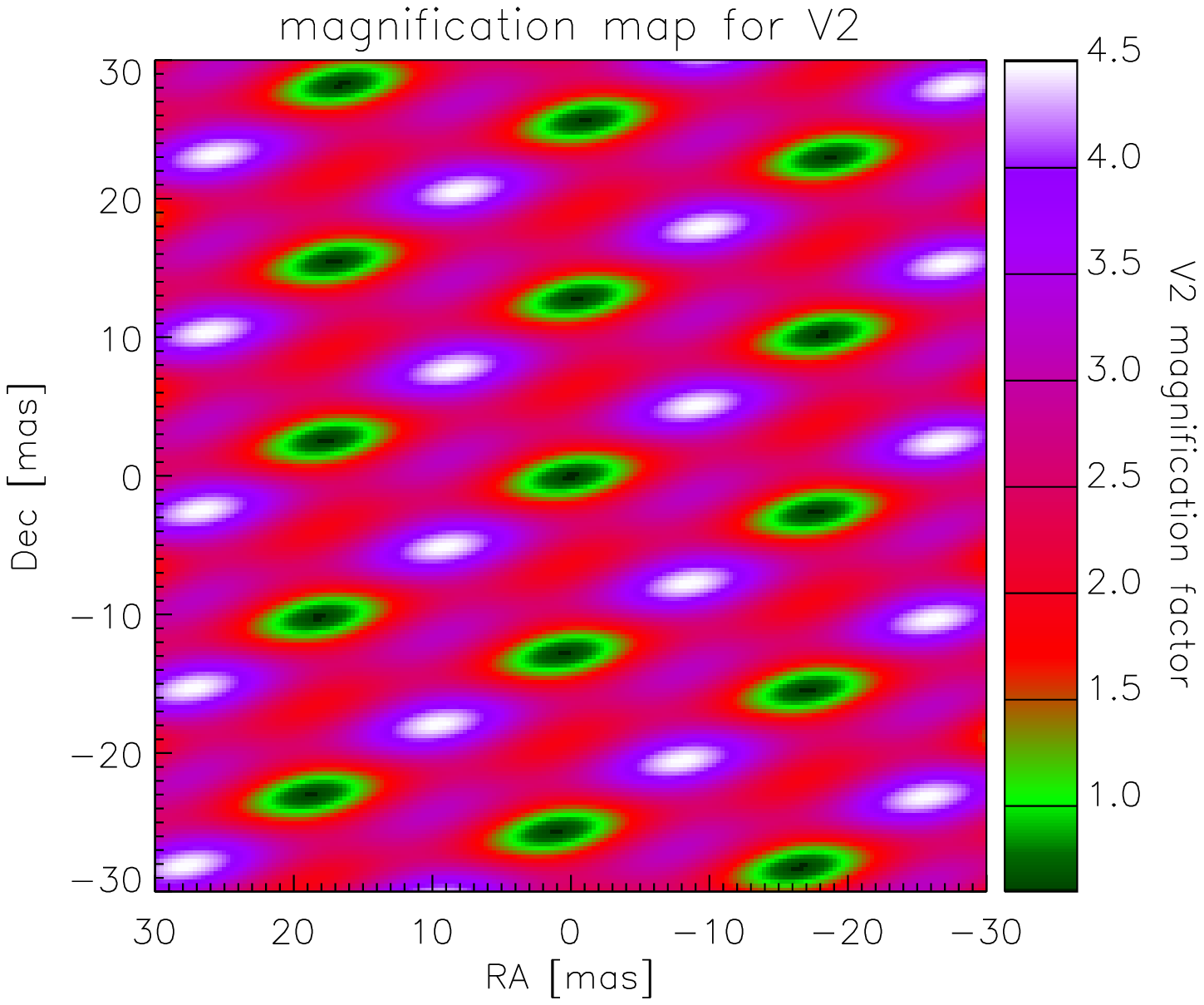}\quad
\includegraphics[scale=0.50]{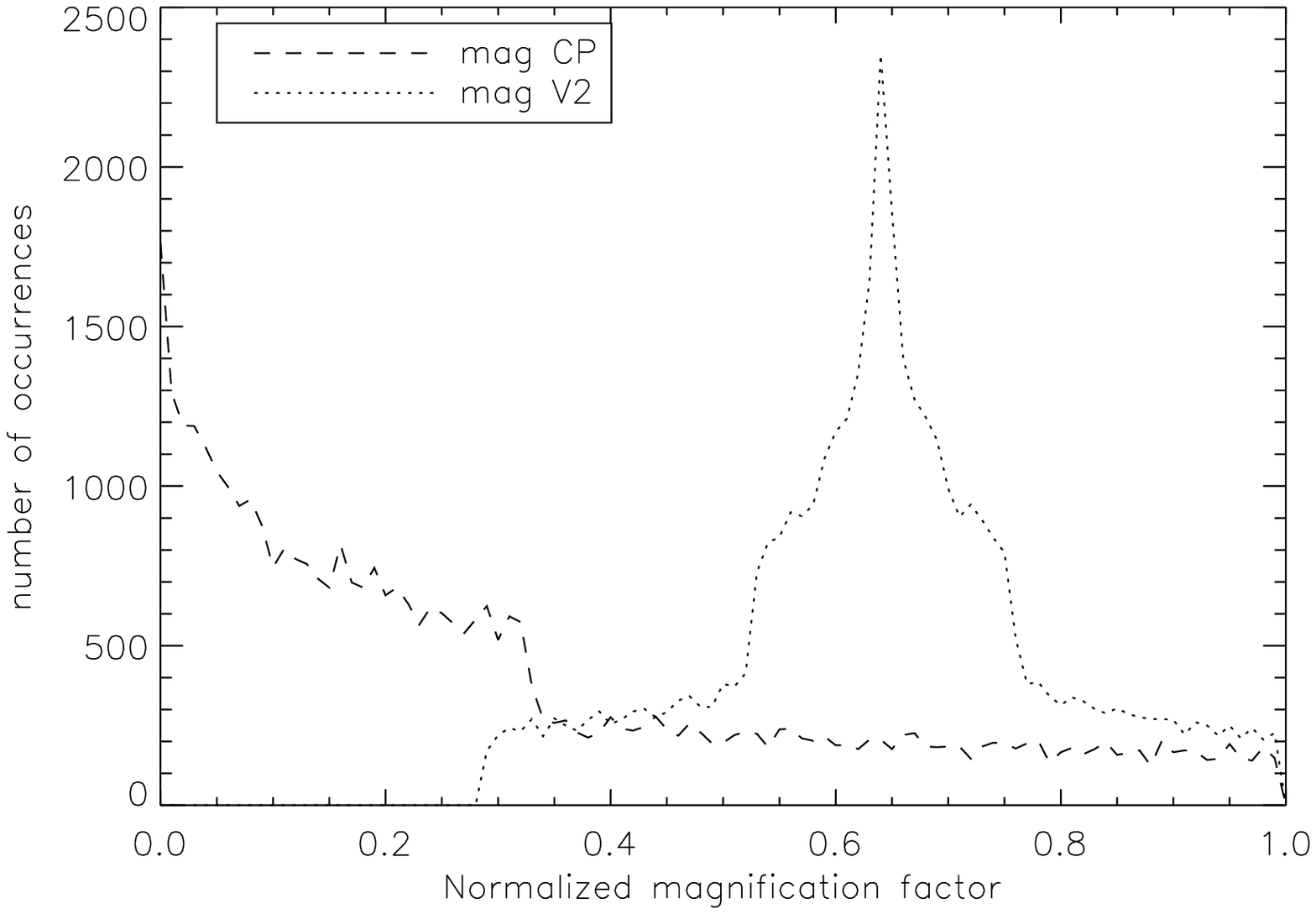}\\
\end{center}
\caption{Magnification map for the closure phase (in absolute value), the square visibility and corresponding histograms for the normalized magnification for both the closure phase and the square visibility, using a linear array (VLTIs tations: D0-G0-H0}
 \label{fig:magmapslin}
\end{figure*}

Thanks to Fig. \ref{fig:sensitivitydet}, we can see that the closure phases and the square visibilities have almost the same median sensitivity. This is somewhat surprising at the closure phases are generally thought to be more sensitive to the presence of a faint companion. Even more surprising, when computing the sensitivity at a 90\% completeness level, we notice that the closure phases are be less sensitive than the square visibilities. To investigate the origin of this unexpected behavior, we compute the so-called ``magnification factor'' ($m$) for the closure phases and squared visibilities on simple three-telescope configurations. The magnification factor is defined as the amplification of the signal of a companion as a function of its position in the field-of-view. It has been formally defined by Le Bouquin \% Absil \cite{lebouquin12} in the case of the closure phase, leading to the following formula:
\begin{equation}
m_{CP} = \sin(\alpha_{12})+\sin(\alpha_{23})-\sin(\alpha_{12}+\alpha_{23}) \; ,
\end{equation}
where 
\begin{equation}
\alpha_{12}=2\pi \frac{\vec{B}_{12}\cdot\vec{\Delta}}{\lambda},\alpha_{23}=2\pi \frac{\vec{B}_{23}\cdot\vec{\Delta}}{\lambda} \; ,
\end{equation}
with $\vec{B}_{ij}$, the projected baseline onto the sky, $\vec{\Delta}$, the apparent binary separation, and $\lambda$, the wavelength of the observation. In the case of the squared visibilities, we used a similar concept to define the magnification map as the combined drop in squared visibilities on the three baselines forming a triangle. Assuming that the primary star is completely unresolved, this boils down to the following formula:
\begin{equation}
m_{V^2} = \sum_{ij}(1-V^2_{ij}) \; ,
\end{equation}
where the baselines, denoted by ``$ij$', can take any of the values 12-23-31 for a three-telescope array. We plot in Fig.\ref{fig:magmaps} and Fig.\ref{fig:magmapslin} the absolute magnification maps for two different baselines configurations (a triangular one and a linear one). We can immediately notice that, in both configurations, the closure phase magnification map comprises much more region where the magnification is close to zero. This is probably related to the fact that the magnification for the closure phase ranges between negative and positive values, while the magnification for the squared visibilities is always positive. The closure phase therefore shows many more ``blind spots'', where the presence of a companion would not show up in the data.

To back up this discussion, we analyze the histograms of the normalized magnification maps for the square visibilities and the closure phases. We notice in Fig.\ref{fig:magmaps} and Fig.\ref{fig:magmapslin} that most of the occurrences for the closure phase are close to zero, which confirm the fact that, even if the closure phases are arguably more robust than the squared visibilities for the detection of a companion, the poor coverage of the field of view limits the confidence level we can have on contrast upper limits. As a conclusion, we see with those plots that taking the square visibilities into account when searching for faint companions is always a good idea, despite the possible presence of false positive in the form of circumstellar disks.

%%-----------------------------------------------------------
\section{CONCLUSION}

In this paper, we have described our method to search for companion with VLTI/PIONIER, and more precisely within the EXOZODI sample. We have proved that the use of the closure phase and the square visibility in a combined way is more effective than using only the closure phases to systematically identify the presence of faint companion and to reject false positive detections, because the square visibility offers a better $u,v$ coverage and is as sensitive as the closure phase to faint companions. Our analysis led to the discovery of five new A-type star binaries: HD4150, HD16555, HD29388, HD202730, and HD224392, although the latter needs further observations to be confirmed.

%%%%%%%%%%%%%%%%%%%%%%%%%%%%%%%%%%%%%%%%%%%%%%%%%%%%%%%%%%%%%
\acknowledgments     %>>>> equivalent to \section*{ACKNOWLEDGMENTS}       
 
The authors thank thank the French National Research Agency (ANR, contract ANR-2010 BLAN-0505-01, EXOZODI) for financial support. L.M.\ acknowledges F.R.S.-FNRS for financial support through a FRIA PhD fellowship. O.A.\ is F.R.S.-FNRS Resarch Associate.This work made use of the Smithsonian/NASA Astrophysics Data System (ADS) and of the Centre de Donn\'ees astronomiques de Strasbourg (CDS).

%%%%%%%%%%%%%%%%%%%%%%%%%%%%%%%%%%%%%%%%%%%%%%%%%%%%%%%%%%%%%
%%%%% References %%%%%

\bibliographystyle{spiebib}   %>>>> makes bibtex use spiebib.bst
\bibliography{procspie}   %>>>> bibliography data in report.bib

\begin{thebibliography}{1}

\bibitem{Ertel14}
{Ertel}, S., {Absil}, O., {Defr\`ere}, D., {Le Bouquin}, J.-B., {Augereau},
  J.-C., {Marion}, L., {Blind}, N., {Bonsor}, A., {Bryden}, G., {Lebreton}, J.,
  and {Milli}, J., ``{A near-infrared interferometric survey of debris disc
  stars. IV. An unbiased sample of 92 southern stars observed at H band with
  VLTI/PIONIER},'' {\em A\&A}  (2014).
\newblock in prep.

\bibitem{Absil11}
{Absil}, O., {Le Bouquin}, J.-B., {Berger}, J.-P., {Lagrange}, A.-M.,
  {Chauvin}, G., {Lazareff}, B., {Zins}, G., {Haguenauer}, P., {Jocou}, L.,
  {Kern}, P., {Millan-Gabet}, R., {Rochat}, S., and {Traub}, W., ``{Searching
  for faint companions with VLTI/PIONIER. I. Method and first results},'' {\em
  A\&A}~{\bf 535},  A68 (Nov. 2011).

\bibitem{2007A&A...475..243D}
{di Folco}, E., {Absil}, O., {Augereau}, J.-C., {M{\'e}rand}, A., {Coud{\'e} du
  Foresto}, V., {Th{\'e}venin}, F., {Defr{\`e}re}, D., {Kervella}, P., {ten
  Brummelaar}, T.~A., {McAlister}, H.~A., {Ridgway}, S.~T., {Sturmann}, J.,
  {Sturmann}, L., and {Turner}, N.~H., ``{A near-infrared interferometric
  survey of debris disk stars. I. Probing the hot dust content around
  {$\epsilon$} Eridani and {$\tau$} Ceti with CHARA/FLUOR},'' {\em A\&A}~{\bf
  475},  243--250 (Nov. 2007).

\bibitem{lebouquin12}
{Le Bouquin}, J.-B. and {Absil}, O., ``{On the sensitivity of closure phases to
  faint companions in optical long baseline interferometry},'' {\em A\&A}~{\bf
  541},  A89 (May 2012).

\end{thebibliography}

\end{document}